\documentclass[UTF8,reprint,twocolumn,aps,pra,amsmath,amssymb,floatfix,superscriptaddress,longbibliography]{revtex4-1}
\usepackage{amsmath}
\usepackage{amssymb}
\usepackage{mathrsfs}
\usepackage{graphicx}
\usepackage{braket}
\usepackage[colorlinks,citecolor=blue]{hyperref}
\usepackage[caption=false]{subfig}
\usepackage[normalem]{ulem}
\usepackage{verbatim}
\usepackage{xcolor}

\usepackage[utf8x]{inputenc}
\usepackage[colorlinks,citecolor=blue]{hyperref}

\def\blue{\textcolor{blue}}
\def\red{\textcolor{red}}

\usepackage{CJK}

\begin{document}

\def\qv{\vec{q}}
\def\red{\textcolor{red}}
\def\blue{\textcolor{blue}}
\def\magenta{\textcolor{magenta}}
\def\apricot{\textcolor{Apricot}}

\def\GJ{\textcolor{black}}
\def\TT{\textcolor{ForestGreen}}
\definecolor{ora}{rgb}{1,0.45,0.2}
\def\LH{\textcolor{black}}

\newcommand{\ad}[1]{\text{ad}_{S_{#1}(t)}}


\title{Non-Hermitian squeezed polarons}

\author{Fang Qin}
\email{qinfang@nus.edu.sg}
\affiliation{Department of Physics, National University of Singapore, Singapore 117542, Singapore}

\author{Ruizhe Shen}
\affiliation{Department of Physics, National University of Singapore, Singapore 117542, Singapore}

\author{Ching Hua Lee}
\email{phylch@nus.edu.sg}
\affiliation{Department of Physics, National University of Singapore, Singapore 117542, Singapore}

\date{\today}
\begin{abstract}
Recent experimental breakthroughs in non-Hermitian ultracold atomic lattices have dangled tantalizing prospects in realizing exotic, hitherto unreported, many-body non-Hermitian quantum phenomena. In this work, we discover and propose an experimental platform for a radically different non-Hermitian phenomenon dubbed polaron squeezing. It is marked by a dipole-like accumulation of fermions arising from an interacting impurity in a background of non-Hermitian reciprocity-breaking hoppings. We computed their spatial density and found that, unlike Hermitian polarons which are symmetrically localized around impurities, non-Hermitian squeezed polarons localize asymmetrically in the direction opposite to conventional non-Hermitian pumping and non-perturbatively modify the entire spectrum, despite having a manifestly local profile. We investigated their time evolution and found that, saliently, they appear almost universally in the long-time steady state, unlike Hermitian polarons which only exist in the ground state. In our numerics, we also found that, unlike well-known topological or skin localized states, squeezed polarons exist in the bulk, independently of boundary conditions. Our findings could inspire the realization of many-body states in ultracold atomic setups, where a squeezed polaron can be readily detected and characterized by imaging the spatial fermionic density.
\end{abstract}

\maketitle

\noindent{\it Introduction.--} 
Rapid recent experimental progress in metamaterial~\cite{feng2011nonreciprocal,shen2016experimental,park2020observation,coulais2017static,zhu2018simultaneous,ghatak2020observation,brandenbourger2019non,gao2021non}, circuit~\cite{helbig2020generalized,hofmann2019chiral,liu2020gain,hofmann2020reciprocal,liu2021non,zou2021observation,stegmaier2021topological,ce2022experimental}, photonic~\cite{xiao2020non,feng2017non,tzuang2014non,zhou2018observation,zhen2015spawning,zeuner2015observation,xiao2020non,miri2019exceptional}, and ultracold atomic~\cite{Jiaming2019gainloss,lapp2019engineering,ren2022chiral,liang2022dynamic,yi2022exceptional,gou2020tunable} 
realizations of non-Hermitian models have made
unconventional features such as exceptional branch points~\cite{Bergholtz2021rmp,lafalce2019robust,kozii2017non,yoshida2019exceptional,yoshida2019symmetry,okugawa2019topological,hodaei2017enhanced,heiss2016circling,heiss2012physics} and non-Hermitian topological windings~\cite{gong2018topological,yoshida2018non,zhang2020correspondence,zhong2018winding,rafi2021topological,yoshida2019mirror,sun2020biorthogonal,xi2021classification}
experimental realities. However, to date, their explorations have mostly been confined to the single-body paradigm, with associated phenomena such as gapped topological transitions~\cite{lee2020unraveling,li2021non}, unconventional criticality~\cite{li2020critical}, negative entanglement entropy~\cite{chang2020entanglement,lee2020exceptional,okuma2021quantum}, and the breakdown of bulk-boundary correspondences~\cite{yao2018edge,xiong2018does,lee2019anatomy,Kunst2018prl,Yokomizo2019prl,Imura2019prb,Song2019BBC,borgnia2020non,li2021quantized,lv2021curving,yang2022designing,jiang2022dimensional,tai2022zoology}. But even more intriguing, many-body phenomena have come within the horizon ever since the very recent experimental breakthroughs in non-Hermitian ultracold atomic setups~\cite{Jiaming2019gainloss,lapp2019engineering,ren2022chiral,liang2022dynamic,yi2022exceptional,gou2020tunable}. The interplay of non-Hermiticity with many-body effects has now become a possibility, as captured by emerging directions such as non-Hermitian many-body localization~\cite{Hamazaki2019prl,Zhai2020prb,suthar2022non,wang2022non}, superfluids~\cite{luo2020skin,Zhang2020prb,Liu2020prb,zhou2019enhanced,Zhou2020pra,pan2021emergent}, and Fermi liquids~\cite{lee2021many,Shen2021arxiv,Pan2020np,kawabata2022many,yamamoto2019theory,alsallom2021fate,zhang2022symmetry,poddubny2022topologically,yoshida2022fate}.

\begin{figure}
\includegraphics[width=1\linewidth]{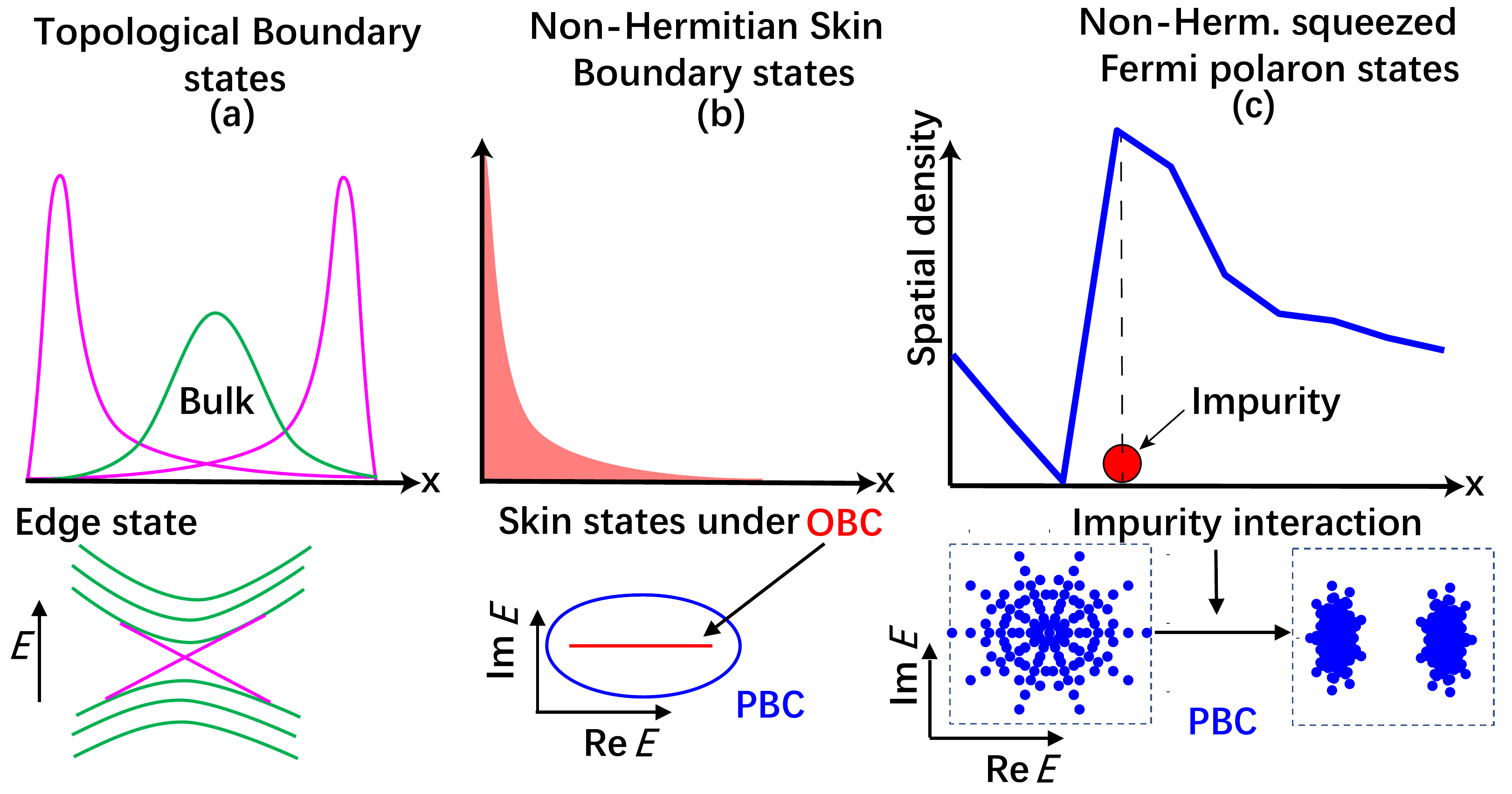}
\caption{Non-Hermitian polaron squeezing is distinct from other mechanisms for localized states, such as (a) topological localization and (b) the non-Hermitian skin effect, both single-body mechanisms requiring open boundaries. By contrast, squeezed polarons (c) are special asymmetric dipole-like accumulations across either side of an interacting impurity. They are many-body dressed states in the bulk, with charge density ``squeezed'' in the opposite direction from non-Hermitian pumping. Illustrative numerics are from $\hat H_\text{min}$ of Eq.~\eqref{Hmin}. 
}
\label{fig:1}
\end{figure}

In this work, through exact diagonalization computations~\cite{Weinberg2017QuSpin,Weinberg2019QuSpin}, we discover a  non-Hermitian many-body phenomenon dubbed ``polaron squeezing'', which is a directional dipole-like accumulation effect arising from the triple interplay of impurity interactions, fermionic statistics, and non-Hermitian flux. In conventional Hermitian settings, polarons are many-body states dressed by the environment-impurity interaction, as observed in ultracold-atom experiments involving both fermions~\cite{Schirotzek2009prl,Zhang2021prl,Kohstall2012nature,Koschorreck2012nature,Cetina2016science,Scazza2017prl,Yan2019prl,oppong2019observation,Ness2020prx} and bosons~\cite{Hu2016prl,Jorgensen2016prl,Yan2020science}.
By providing a unique angle for understanding strong interactions in solid-state and cold-atom systems, they are valuable probes for detecting quantum phase transitions in interacting topological settings~\cite{hu2013universal,lee2014lattice,lee2015geometric,grusdt2016interferometric,Camacho2019prb,Pimenov2021prb,qin2019polaron}. 

Going beyond well-understood Hermitian polarons~\cite{Chevy2006pra,Lobo2006prl,Combescot2007prl,Bruun2010prl,Mathy2011prl,zollner2011polarons,parish2011polaron,levinsen2012p,Yi2012prl,Parish2013pra,Yi2015pra,nishida2015polaronic,parish2016quantum,Camacho2018prl,qin2019polaron,cui2020fermi,pessoa2021finite,ardila2021dynamical,seetharam2021dynamical,seetharam2021quantum,hu2022crossover,qin2022light,qin2022phase,qin2020theory}, we found that, with non-Hermiticity and flux, polarons can acquire interesting aggregate behavior, with chiral delocalizing tendencies competing with impurity localization, in a way distinct from non-interacting impurities under the non-Hermitian skin effect (NHSE)~\cite{li2021impurity}. Fermion degeneracy pressure introduces another level of intrigue by enforcing a special type of equilibrium among these competing influences. The result is a unique real-space ``squeezed'' fermionic density profile that, as we show, can be feasibly imaged in a realistic ultracold atomic demonstration.

Arising from predominantly many-body mechanisms, squeezed polaron states are distinct from other existing well-known types of robust states in related physical settings. Chiral topological states~\cite{Hasan2010rmp,Chiu2016rmp} [Fig.~\ref{fig:1}(a)], for instance, are edge localized and asymmetrically propagating, but they originate from nontrivial Chern topology, which is already completely well-defined in the single-particle context. Non-Hermitian boundary-localized skin states~\cite{yao2018edge,xiong2018does,lee2019anatomy,Kunst2018prl,Yokomizo2019prl,Imura2019prb,Song2019BBC} [Fig.~\ref{fig:1}(b)] are also essentially single-particle phenomena, with their robustness stemming from the directed non-Hermitian ``pumping'' in non-reciprocal lattices. In contrast, squeezed polarons [Fig.~\ref{fig:1}(c)] are bona fide many-body states localized beside an impurity \emph{interacting} with the fermions, and they can exist without nontrivial topology or physical boundaries. Due to their many-body nature, squeezed polarons also exhibit spatial profiles that are very different from those of topological or skin states.

\noindent{\it Squeezed polarons from interactions and non-reciprocal gain and loss.--} To understand the primary mechanism behind polaron squeezing, we first examine a minimal toy model $\hat{H}_\text{min}$ where fermions interact with a single impurity with strength $g$ and hop asymmetrically with amplitudes $e^{\pm \alpha}$ around a ring with circumference $L$ that gives periodic boundary conditions (PBCs):
\begin{equation}
\hat{H}_\text{min}\!=\!g\hat b^\dagger_{x_0}\hat b_{x_0}\hat c^\dagger_{x_0}\hat c_{x_0}\!+\!\sum_{x}(e^{\alpha}\hat c_x^\dagger\hat c_{x+1}\!+\!e^{-\alpha}\hat c_{x+1}^\dagger\hat c_x).
\label{Hmin}
\end{equation} 
Here $\hat c$ and $\hat b$ are respectively the second-quantized operators for the fermions and the impurity, which is fixed at an arbitrary site $x_0$. They experience a density-density interaction of strength $g$. The fermions also experience asymmetric hoppings $e^{\pm \alpha}$, which are the simplest possible terms that represent the simultaneous breaking of Hermiticity and reciprocity~\footnote{For a more practical implementation involving physical flux, see the more realistic cold-atom Hamiltonian presented later}. Importantly, due to the PBCs, these asymmetric hoppings cannot be ``gauged away'' as in conventional literature on the boundary accumulation of non-Hermitian skin states~\cite{yao2018edge,lee2019anatomy}. This independence from boundary accumulation is the first hint of the fundamental distinction between squeezed polarons and topological as well as skin states.

Squeezed polarons arise when the two parameters $g$ and $\alpha$ of $\hat{H}_\text{min}$ are both nonzero and sufficiently large. To elucidate their behavior, we turn on $g$ and $\alpha$ under PBCs and observe how that affects the energy spectrum and long-time steady-state spatial density
\begin{align}\label{rho}
\rho(x)\equiv\lim_{t\to\infty}\langle\psi^{R}(t)|\hat{c}_{x}^\dagger\hat{c}_{x}|\psi^{R}(t)\rangle.
\end{align} 
Here 
$|\psi^{R}(t)\rangle=e^{-i\hat{H}t}\left|\psi^{R}(0)\right\rangle/\left\|e^{-i\hat{H}t}\left|\psi^{R}(0)\right\rangle\right\|$ is the normalized $N$-fermion right eigenstate that has time evolved from a specified initial state $\left|\psi^{R}(0)\right\rangle$. This evolution is taken over a sufficiently long time $t$, such that the spatial density approaches a steady spatial configuration. 

\begin{figure}
\includegraphics[width=1\linewidth]{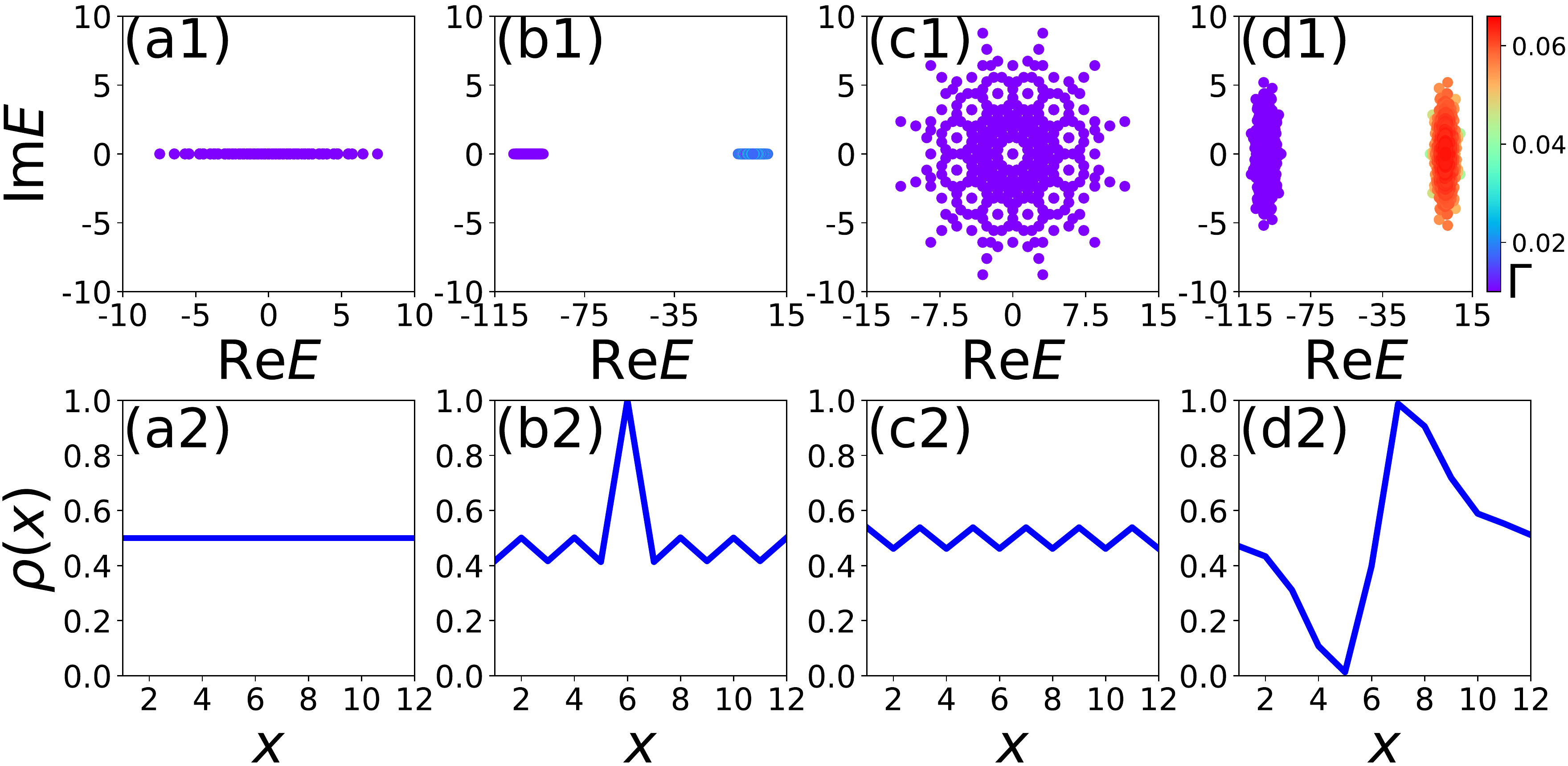}\\
\includegraphics[width=1\linewidth]{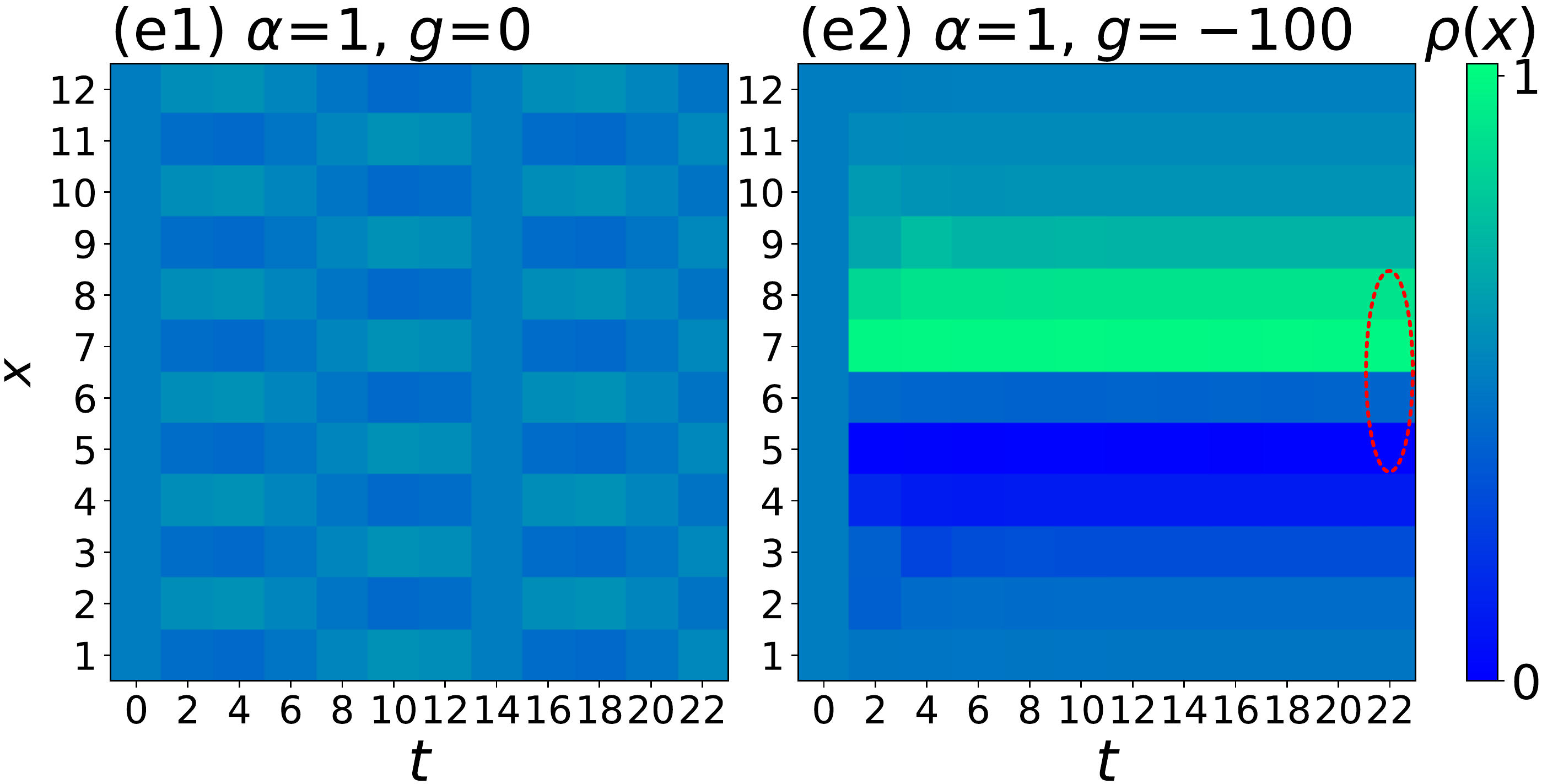}
\caption{PBC spectrum $E$ [panels (a1)-(d1)] and spatial density $\rho(x)$ [panels (a2)-(d2)] for $\hat H_\text{min}$ [Eq.~\eqref{Hmin}] with different non-reciprocities $\alpha$ and impurity interaction strengths $g$: (a1,a2) $\alpha=g=0$; (b1,b2) $\alpha=0$, $g=-100$; (c1,c2) $\alpha=1$, $g=0$; and (d1,d2) $\alpha=1$, $g=-100$. Energy eigenstates $|\psi\rangle$ are colored by their squeezing asymmetric parameter $\Gamma$, which captures polaron squeezing: $\Gamma$ is large only with both non-reciprocity and impurity interaction (d1). While the spatial polaron density $\rho(x)$ is symmetrically peaked about the impurity at $x_0=6$ when Hermitian (b2), it is asymmetrically squeezed in the non-Hermitian case (d2). (e1,e2) Dynamics of the spatial density for (e1) $\alpha=1$, $g=0$ and for (e2) $\alpha=1$, $g=-100$, with the dipole-like asymmetric profile as circled. All computations are with $N=6$ fermions in $L=12$ sites. The initial state $\left|\psi^{R}(0)\right\rangle$ is the ground state for $\alpha=0$ and is $(\left|101010101010\right\rangle+\left|010101010101\right\rangle)/\sqrt{2}$ for $\alpha=1$~\cite{supp}.}
\label{fig:2}
\end{figure}

When $\alpha=g=0$ [Figs.~\ref{fig:2}(a1) and \ref{fig:2}(a2)], we trivially have Hermitian nearest-neighbor hoppings with a real gapless spectrum. Due to translation invariance from PBCs, $\rho(x)=0.5$ everywhere. Turning on the impurity interaction such that $\alpha=0$ and $g=-100$ [Figs.~\ref{fig:2}(b1) and \ref{fig:2}(b2)], we realize a minimal Hermitian polaron bound state, with $\rho(x)$ peaking at the impurity $x_0$. It is the polaron bound by the gap which opens up. When we turn on the non-Hermiticity and non-reciprocity instead of the interaction, such that $\alpha=1$ and $g=0$ [Figs.~\ref{fig:2}(c1) and \ref{fig:2}(c2)], $\rho(x)$ is oscillating around $0.5$, and the spectrum becomes complex with  $L$ star-like spikes~\cite{supp}. Note that it is not a superposition of the energies of $N$ Hatano-Nelson chains~\cite{hatano1996localization,hatano1997vortex,hatano1998non-Hermitian}, since Pauli exclusion constrains certain asymmetric hoppings. 

Finally, turning on both the interaction and non-Hermiticity such that $\alpha=1$ and $g=-100$ [Figs.~\ref{fig:2}(d1) and \ref{fig:2}(d2)], we observe a peculiar state with an asymmetric profile $\rho(x)$ around the impurity at $x_0+1$, which we name a ``squeezed polaron''. The density to the left of $x_0+1=7$ (sites 5 and 6) appears to be ``squeezed'' towards the right (sites 8 and 9) by the impurity interaction, even though, naively, we would have expected the asymmetric $e^{\pm \alpha}\hat{c}_x\hat{c}_{x\pm 1}$ couplings to pump the states from right to left instead. Notably, the $\rho(x)$ peak is not exponentially high like topological or non-Hermitian skin states, but instead resembles a finite local dipole within the Fermi sea. The spectrum is complex, and a gap separates two almost identical star-like ``bands'', the one with negative $\text{Re}(E)$ containing states bounded by the attractive ($g<0$) impurity interaction. 

\begin{figure}
\includegraphics[width=1\linewidth]{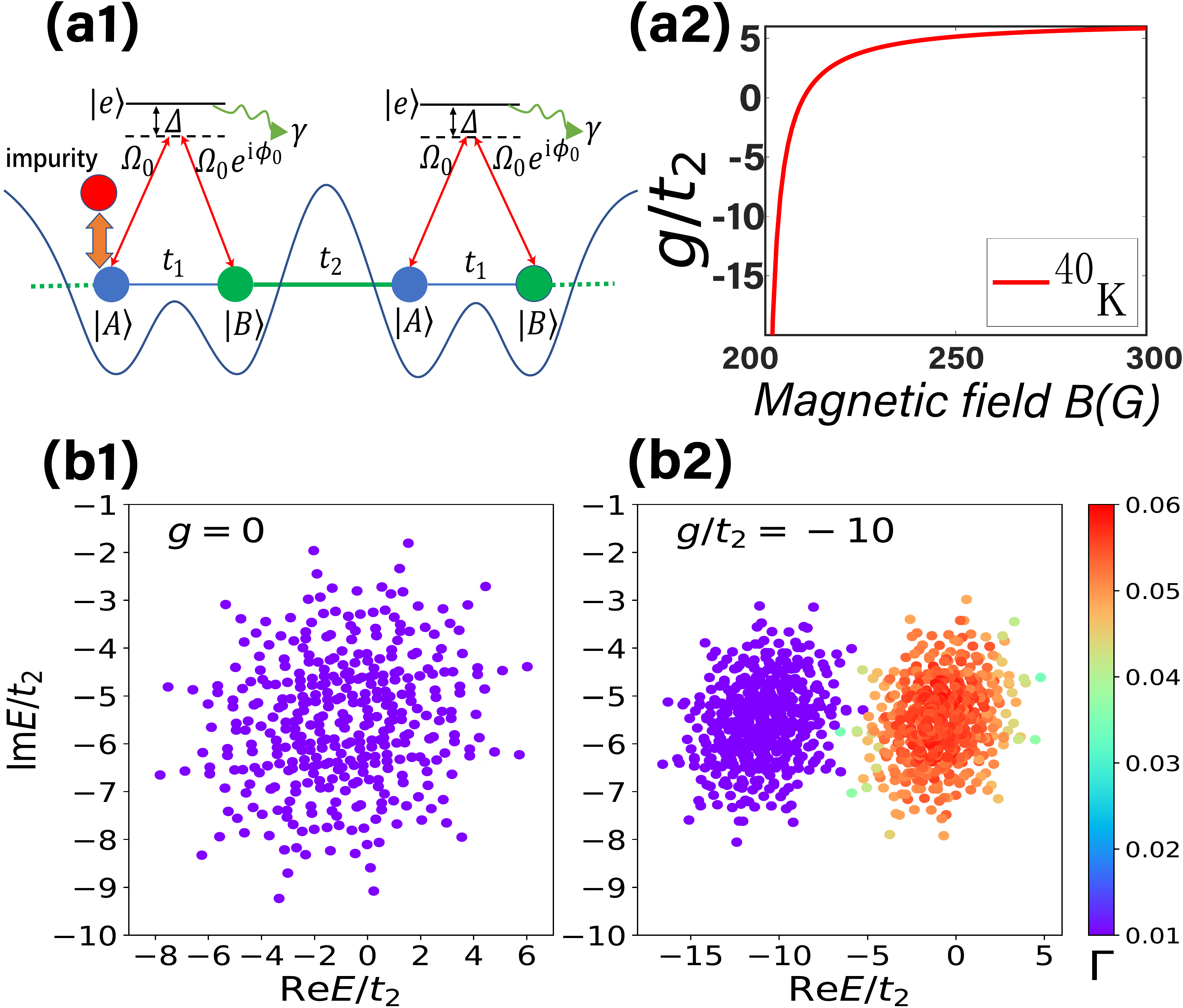}
\caption{(a1) Our effective interacting Hamiltonian for squeezed polarons [Eq.~\eqref{eq:H}] is based on a two-photon dissipative Raman process~\cite{Zhou2020pra,gou2020tunable} with impurity interactions $g$ from Feshbach resonance. The Rabi frequency is $\Omega_0$ between hyperfine ground states $|A\rangle$ and $|B\rangle$ and the excited state $|e\rangle$ for $^{40}$K atoms; a phase difference $\phi_0$ introduces non-reciprocity. $\Delta$ is the single-photon detuning of the excited state $|e\rangle$, whose non-Hermitian decay rate $\gamma$ can be laser controlled.
(a2) The impurity interaction $g$ is highly tunable through the magnetic field, with parameters given by Refs.~\cite{Liao2010nature,Regal2004prl,Chin2010rmp}. 
(b1,b2) $g$ non-perturbatively modifies the PBC spectrum $E$ at half-filling $N=6$ and $2L=12$, such that all states $|\psi\rangle$ become squeezed with elevated squeezing asymmetry $\Gamma$ when $|g|\neq 0$.
Here, $t_2=(2\pi)\times 1000$ Hz~\cite{Atala2013np} sets the energy scale.}
\label{fig:3}
\end{figure}

Interestingly, even though the impurity interaction acts locally, its presence affects the \emph{entire} spectrum [Fig.~\ref{fig:2}(d1)], not just states localized around the impurity. This is most saliently revealed through the squeezing asymmetry parameter $\Gamma$ of a given $N$-fermion state $|\psi\rangle$, which we define as
\begin{align}
\Gamma\equiv\sum_{x=1}^{L}(x-x_0-1)e^{-(x-x_0-1)^{2}}|\langle\psi|\hat n_x|\psi\rangle|^{2}/N.
\label{Gammaj}
\end{align} 
Containing the derivative of a Gaussian kernel, it measures the extent of asymmetric state localization around the the impurity neighbor $x_0+1$, unlike the more commonly used inverse participation ratio (IPR) parameter~\cite{hikami1986localization}, which is agnostic to the localization asymmetry and position. In particular, it distinguishes our squeezed polarons from ordinary polarons in Hermitian settings, which are symmetric about the impurity. As a reference, a profile with a perfectly localized surplus particle on each side has $\Gamma=2/(Ne)\approx 0.74/N$, which is just slightly higher than the $\Gamma$ of the eigenstates with polaron squeezing behavior [Figs.~\ref{fig:2}(d1) and \ref{fig:3}(b2)].  
This also implies that the squeezed polaron is distributed across \emph{all} bound states, and not particular ground states as with ordinary polarons. Physically, this is because the impurity interaction has become effectively non-local in the background of non-reciprocal gain and loss pumping; but contrary to a simple pumping of states, what we observe is an interaction-facilitated ``squeezing'' in the opposite direction that results in a dipole-like density profile. Herein lies an important physical distinction between Hermitian polarons and non-Hermitian squeezed polarons - while squeezed polaron asymmetry can be observed in the long-time steady state evolved from most generic initial states~\footnote{This is because a randomly given initial state would most likely overlap with some of the many eigenstates in the right (reddish) cluster of Fig.~2(d1).} [Fig.~\ref{fig:2}(e2)], Hermitian (symmetric) polaron localization only exists for the ground state (Fig.~S8 of~\cite{supp}).

\noindent{\it Ultracold atomic model for observing squeezed polarons.--} Having discussed the essential though simplified mechanism behind squeezed polarons, we now turn to a more realistic setup without asymmetric physical couplings and that can be feasibly implemented in an ultracold atomic setup. 

The key ingredients for squeezed polarons are (i) impurity interaction, (ii) non-reciprocity, and (iii) loss. To incorporate them all, we consider a one-dimensional fermionic array of $N$ fermionic $^{40}$K atoms, with the majority being spin $\uparrow$ and the minority being spin $\downarrow$ impurities. 

To implement the impurity interaction (i), we apply an external magnetic field $B$ that causes the atoms to experience a strong Feshbach resonance~\cite{Olshanii1998prl,Olshanii2003prl,Chin2010rmp,Qin20201pra,Qin20202pra,Chin2010rmp,Regal2004prl,Regal2003prl,Qin2016pra,Qin2017epjd} that corresponds to a density-density $S$-wave interaction, 
\begin{equation}
\hat H_\text{int}=g\hat{n}_{x_0,s}^{(b)}\hat{n}_{x_0,s},
\end{equation}
between unlike (majority and impurity) spins, where $\hat{n}_{x_0,s}^{(b)}=\hat{b}_{x_0,s}^{\dagger}\hat{b}_{x_0,s}$ is the number density operator of the spin-$\downarrow$ impurity atom, which is situated at site $s$ of the $x_0$th unit cell, and $\hat{n}_{x_0,s}=\hat{c}_{x_0,s}^{\dagger}\hat{c}_{x_0,s}$ is the corresponding density operator of spin-$\uparrow$ majority atoms at the same position. The interaction strength $g\sim g_0(B-B_c)^{-1}$ 
becomes very strong and saturates at a large value near a resonant magnetic field~\footnote{In practice, at very strong resonance, the scattering length diverges and a quantum halo state is formed instead. } $B=B_c$, as numerically computed~\cite{Olshanii1998prl,Olshanii2003prl,Chin2010rmp,Qin20201pra,Qin20202pra} and plotted in Fig.~\ref{fig:3}(a2), and can be tuned to any desired strength between $-1000t_2$ and $\approx 1500t_2$ by appropriately adjusting the field strength~\cite{supp}.

A non-reciprocal lattice with loss [(ii) and (iii)] can be achieved by coupling a two-photon dissipative Raman process to the discrete hyperfine ground states $|A\rangle$ and $|B\rangle$~\cite{liang2022dynamic,Zhou2020pra,gou2020tunable,Lin2009prl,Lin2011nature,Zhang2012prl,Wang2012prl,Cheuk2012prl,Qu2013pra,li2019topology} of each degenerate $^{40}$K atom and subjecting the atoms to a strong periodic optical potential~\cite{gou2020tunable,Atala2013np,Folling2007nature}, as schematically illustrated in Fig.~\ref{fig:3}(a1). Non-reciprocity is introduced through the phase difference $ \phi_0$ between the optical fields exciting each hyperfine state; for maximum time-reversal breaking, we set $\phi_0=\pi/2$. By adiabatically eliminating~\footnote{See Supplemental Material for adiabatically elimination, which includes~\cite{Mila2010arxiv,Mila2011book,Sakurai1994book}} the excited state $|e\rangle$, one obtains an effective spin-orbit coupling in the pseudospin basis of $|A\rangle$ and $|B\rangle$. If the excited state additionally experiences laser-induced decay of rate $\gamma$, the coupling becomes effectively complex~\footnote{See Supplemental Material for the derivation of the effective model, which includes~\cite{Zhou2020pra,Zhou2021epl,Manzano_2020,pan2021point-gap,kawabata2022entanglement}}, leading to an effective tight-binding Hamiltonian ($\hbar=1$),
\begin{align}
\hat{H}&=\sum_{x}^{L}\left[\left(t_{1}+\tilde{\gamma}\right)\hat{c}_{x,A}^{\dagger}\hat{c}_{x,B} + \left(t_{1}-\tilde{\gamma}\right)\hat{c}_{x,B}^{\dagger}\hat{c}_{x,A}\right. \nonumber\\
&~~+ \left. t_{2}(\hat{c}_{x+1,A}^{\dagger}\hat{c}_{x,B} + {\rm H.c.})\right]+ g\hat{n}_{x_0,s}^{(b)}\hat{n}_{x_0,s} - i\tilde{\gamma}\sum_{x,s}^{L}\hat n_{x,s}  , \label{eq:H}
\end{align} 
where $t_1$ and $t_2$ depend on the optical potential~\cite{supp} and
\begin{align}
\tilde{\gamma} = \frac{\Omega_{0}^{2}}{\gamma+i\Delta}
\end{align} 
is the effective decay rate, where $\Omega_0$ and $\Delta$ are the single-photon Rabi frequency and detuning respectively. The effective intra-cell hoppings $t_1\pm \tilde \gamma$ have become asymmetric and complex due to the combination of the reciprocity breaking and dissipation, even though the physical optical lattice couplings are all symmetric~\footnote{See Supplemental Material for mapping to tight-binding model, which includes~\cite{Liu2013prl,Liu2014prl,Pan2015prl,Fan2018pra}}. 

Before presenting the numerical results on this model, we briefly outline the experimental specifications for the parameters used. First, a tiny fraction of spin-$\downarrow$ ``impurity'' atoms can be created by exciting a spin-polarized (spin-$\uparrow$) cloud of $^{40}$K atoms in the two lowest hyperfine states~\cite{Schirotzek2009prl,Koschorreck2012nature} via a two-photon Landau-Zener sweep and subsequently cooling it down. The resultant Fermi gas is then loaded onto a one-dimensional optical superlattice potential $V(x)$~\cite{Atala2013np,Folling2007nature}, which is formed by superimposing two standing optical lasers with wavelengths $\lambda_2=767$ nm (short lattice) and $\lambda_1=2\lambda_2$ (long lattice)~\cite{Atala2013np,Folling2007nature}, such that $V(x)=V_{1}\sin^{2}(k_{1}x+\phi_{0}/2) + V_{2}\sin^{2}(k_{2}x+\pi/2)$, where $k_{1}=2\pi/\lambda_{1}$ and $k_{2}=2k_{1}$, corresponding to a unit cell of size $d=767$ nm~\cite{Atala2013np}. By adjusting the laser amplitudes $V_{1}$ and $V_{2}$, the effective lattice couplings $t_1$ and $t_2$ can be tuned within $(2\pi)\times[60,1000]$ Hz~\cite{Atala2013np,supp}; in this work, we set $t_2=(2\pi)\times1000$ Hz~\cite{Atala2013np} as the reference energy unit, and we fix $t_{1}/t_{2}=1$. For the two-photon dissipative Raman process, we used $\Omega_{0}=(2\pi)\times0.03$ MHz~\cite{Fu2013pra} and $\Delta=(2\pi)\times1$ MHz for the single photon Rabi frequency and detuning, and we fix the adjustable decay rate from the excited state $|e\rangle$ as $\gamma=(2\pi)\times6$ MHz~\cite{Fu2013pra,Jie2017pra,Qin2018pra}, such that the effective decay rate takes the value $\tilde{\gamma}\sim(0.92-0.15i)t_2=(2\pi)\times(0.92-0.15i)$ kHz~\cite{supp}. We fix the impurity at site $s=A$ of the $x_0$th cell. In all, lasers are employed for various distinct purposes: defining the optical lattice potential, sweeping to produce the impurities, and Raman transitions and laser-induced dissipation as shown in Fig.~\ref{fig:3}(a1).

As evident in Figs.~\ref{fig:3}(b1) and (b2), the effective ultracold-atomic Hamiltonian $\hat H$ of Eq.~\eqref{eq:H} captures the essential polaron behavior already present in the minimal single-component model $\hat H_\text{min}$ of Eq.~\eqref{Hmin}, with qualitatively similar many-body spectra. At a relatively modest interaction strength of $g=-10t_2$, corresponding to $B\approx 203.5$G, the spectrum separates into two distinct ``bands'', both of which correspond to squeezed eigenstates. Their squeezed profile $\rho(x)$ (Fig.~S6(a1)-S6(d1)~\cite{supp}) also retains the characteristic asymmetrically squeezed shape, although it also exhibits step-like kinks due to the symmetry breaking from odd (even) $|A\rangle$ ($|B\rangle$) sites~\cite{supp}.

\begin{figure}[h]
\includegraphics[width=1\linewidth]{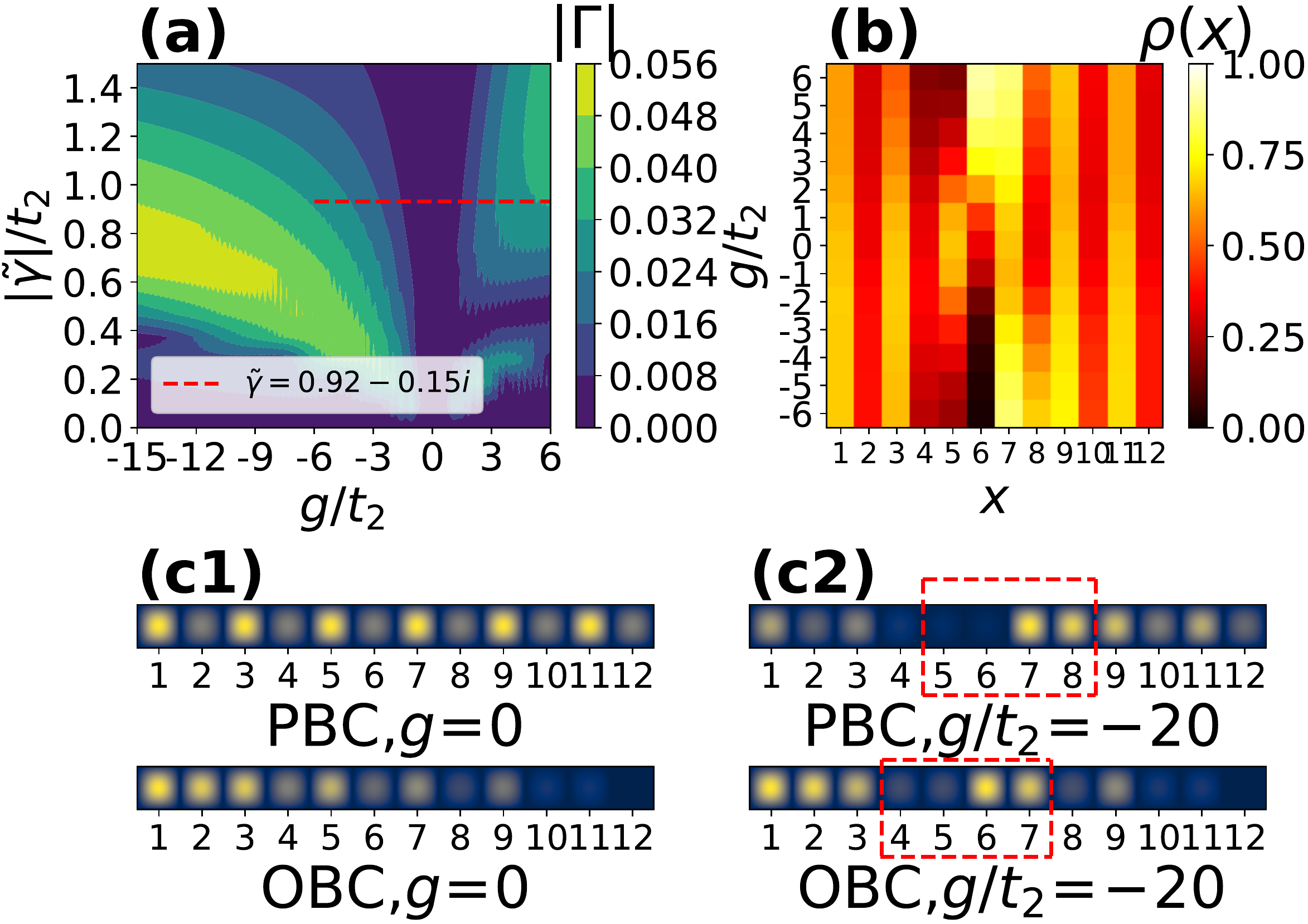}
\includegraphics[width=1\linewidth]{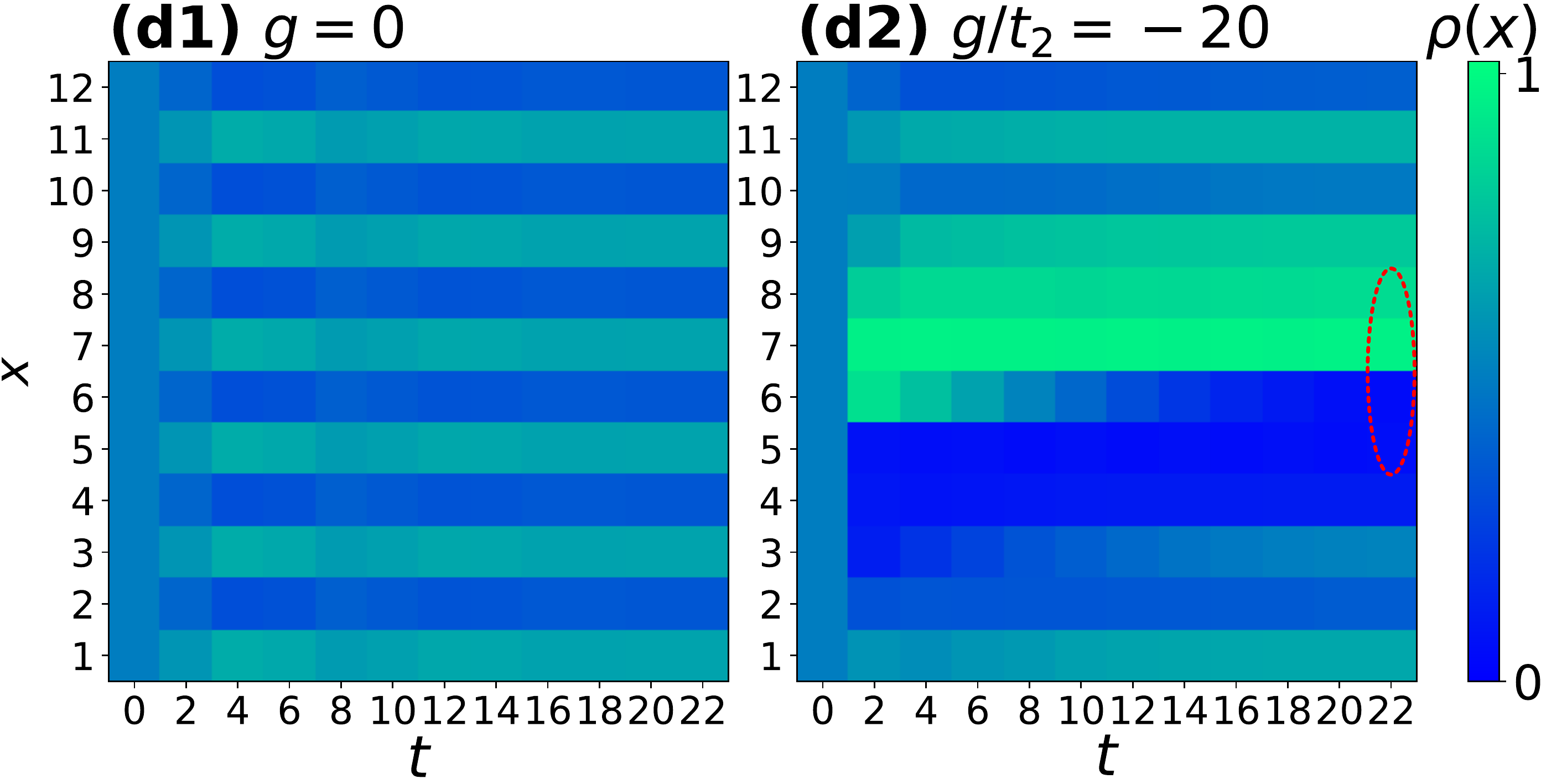}
\caption{
(a) Squeezing expectation $|\Gamma|$ in the $g$-$|\tilde{\gamma}|$ parameter space; note the vanishing squeezing at zero impurity interaction $g$ and the enhanced polaron squeezing at large effective decay rate $|\tilde \gamma|$ and very attractive ($g<0$) or repulsive ($g>0$) interactions. Here, $\Gamma$ is computed from the long-time steady state evolved from the initial state $\left|\psi^{R}(0)\right\rangle=(\left|101010101010\right\rangle+\left|010101010101\right\rangle)/\sqrt{2}$.
(b) Spatial density $\rho(x)$ as a function of $x$ and $g$ under PBCs. Note the very pronounced asymmetric profile across the impurity position $x_0=6$, particularly in the repulsive ($g>0$) case where $\rho(x_0)$ is strongly localized. Data are plotted at $\tilde{\gamma}/t_{2}=0.92-0.15i$, as indicated by the dashed red line in panel (a).
(c1,c2) Simulated spatial density measurements. The density around the impurity ($x_0+1=7$) exhibits almost identical asymmetric profiles of squeezed polaron states (red dashed box) regardless of OBCs or PBCs, as long as the impurity interaction $g$ is nonzero, with a slight shift from sublattice effects. (d1,d2) Evolution of spatial densities from the initial state $\left|\psi^{R}(0)\right\rangle$ under PBCs, with an asymmetric steady-state squeezed polaron profile for (d2) $g\neq 0$. We used $\tilde{\gamma}/t_2=0.92-0.15i$, $t_1/t_2=1$, and $N=6$ fermions in $2L=12$ sites for all subfigures, and we used $g=0$ for panels (c1) and (d1) and $g/t_2=-20$ for panels (c2) and (d2). 
}
\label{fig:4}
\end{figure}

\noindent{\it Attractive vs. repulsive polaron squeezing.--} Figure~\ref{fig:4}(a) shows the squeezing expectation $|\Gamma|$~\footnote{Here $\Gamma$ is contributed by the $2L$ sites of the cold-atom setup.} in the parameter space of $g/t_2$, the normalized impurity interaction strength, and $|\tilde \gamma|/t_2$, the normalized effects of reciprocity and dissipation. For Hermitian scenarios with $|\tilde{\gamma}|=0$ and $g\neq 0$, we indeed have vanishing $\Gamma$, as expected from ordinary polarons with symmetric impurity localizations. In general, the squeezing expectation $|\Gamma|$ increases with larger $|\tilde \gamma|$ or $|g|$, consistent with the intuition that polaron squeezing requires the combined interplay of interactions, non-reciprocity and non-Hermiticity.

However, Fig.~\ref{fig:4}(a) also shows a marked asymmetry between attractive ($g<0$) and repulsive ($g>0$) squeezed polarons. A stronger interaction is required to produce an attractive squeezed polaron, relative to a comparably squeezed repulsive polaron. The reason behind this is clear from the plot of spatial density $\rho(x)$ vs $g/t_2$ [see Fig.~\ref{fig:4}(b)], evaluated at the value of $\tilde\gamma_2=0.92-0.15i$ used in Fig.~\ref{fig:3}. For attractive polarons with $g<0$, $\rho(x)$ is strongly localized at $x_0+1=7$ next to the impurity, leaving a ``hole'' at the impurity. However, for repulsive polarons with $g>0$, $\rho(x)$ is strongly localized at the impurity position $x_0=6$. That said, for both attractive and repulsive cases, the asymmetry in the $\rho(x)$ profile is still strongly contributed by the $\tilde\gamma$ asymmetry. In all, repulsive polarons generally possess a stronger combined ``dipole'' moment and hence larger $\Gamma$.

\noindent{\it Independence from boundary conditions.--} While we have emphasized that squeezed polarons, unlike skin or topological states, are interacting phenomena and not boundary phenomena, actual experimental lattices are usually bounded~\footnote{Actual experimental lattices are usually bounded except in circuit implementations~\cite{ningyuan2015time,lee2018topolectrical}, where the wire connectivity is not tied to their physical embedding.}. Fortunately, that is not a practical obstacle, because squeezed polarons are largely unaffected by boundary conditions, be they OBCs or PBCs. Shown in Fig.~\ref{fig:4}(c), there are simulated spatial density of states for $\rho(x)$ measurements with $N=6$ fermions on $2L=12$ sites and the impurity at $x_0=6$. Without interactions, i.e., $g=0$ (left), we observe skin boundary accumulation under OBCs but not PBCs. However, squeezed polaron physics dominates in the bulk when the impurity interaction is turned on (right). For both PBCs and OBCs, approximately equal Fermi polaron squeezing (red highlighted) counteracts the background skin accumulation, if any. Despite finite-size effects, polaron squeezing is evidently a robust non-Hermitian interaction effect distinguishable from competing single-body effects away from the boundaries.

\noindent{\it Discussion.--} With very recent breakthroughs in non-Hermitian cold-atom experiments~\cite{Jiaming2019gainloss,lapp2019engineering,ren2022chiral,liang2022dynamic,yi2022exceptional,gou2020tunable}, the physical realization of interacting many-body effects is closer to becoming a practical reality even in non-Hermitian settings. We are hopeful that, through our proposal, squeezed polarons can be measured in the near future, thereby realizing a many-body form of emergent non-locality distinct from non-Hermitian skin sensitivity. 

While a squeezed polaron manifests as a local dipole-like density asymmetry, not colossal exponential state localization, it non-perturbatively splits the \emph{entire} spectrum into two halves, a fascinating demonstration of how particle statistics can help encode non-local effects despite seemingly local density effects. While we have largely demonstrated polaron squeezing with PBCs, it occurs independently of boundaries and remains robust in realistic experimental setups subject to OBCs.

\begin{acknowledgements}
\noindent{\it Acknowledgements.--} We thank Shun-Yao Zhang and Hong-Ze Xu for helpful discussions. 
The models are numerically calculated with QuSpin~\cite{Weinberg2017QuSpin,Weinberg2019QuSpin}.
This work is supported by the Singapore National Research Foundation (Grant No.~NRF2021-QEP2-02-P09).
F.Q. acknowledges support from the National Natural Science Foundation of China (Grant No.~11404106), and the project funded by the China Postdoctoral Science Foundation (Grants No.~2019M662150 and No.~2020T130635) before he joined NUS.
\end{acknowledgements}

\bibliographystyle{ieeetr} 
\bibliography{references}


\clearpage
\appendix
\onecolumngrid
\setcounter{equation}{0}
\setcounter{figure}{0}
\setcounter{table}{0}
\setcounter{page}{1}
\renewcommand{\theequation}{S\arabic{equation}}
\renewcommand{\thefigure}{S\arabic{figure}}
\renewcommand{\thesection}{S\Roman{section}}
\renewcommand{\thepage}{S\arabic{page}}
\renewcommand*{\citenumfont}[1]{S#1}
\renewcommand*{\bibnumfmt}[1]{[S#1]}

\section*{\normalsize Supplementary Material for ``Non-Hermitian Squeezed Polarons''}
{\small This appendix contains the following material arranged by sections:\\
	\begin{enumerate}
		\item Derivation of the effective model.
		\item Additional data on the energy spectra and spatial density.
	\end{enumerate}

\section{Derivation of the effective model}\label{1}
Here, we rigorously derive the tight-binding model (Eq.~(5) of the main text) representing our ultracold atomic setup for non-Hermitian squeezed polarons, beginning with its physical Hamiltonian. Starting from the Schr{\"o}dinger's picture formulation for a single unit cell, we switch to the interaction picture to extract the rapidly oscillating phases that can be eliminated via the rotating-wave approximation. Then we go to the rotating frame to eliminate the remaining explicit time dependence. We next treat the most subtle aspect, the dissipative mechanism on the excited state, via the Lindblad master equation, and from it derive the effective non-Hermitian Hamiltonian by adiabatically eliminating the excited state. Incidentally, the latter also agrees with a simple perturbative approach with a phenomenological dissipative term. Next, we consider a 1D array of such setups, and derive a tight-binding Hamiltonian that serves as the non-interacting background of our non-Hermitian squeezed polaron setup.

\subsection{Schr{\"o}dinger picture}

\noindent We first ignore any dissipative effects, and derive the effective static Hamiltonian that encapsulates the optical driving. In the Schr{\"o}dinger picture, our Hamiltonian corresponding to the configuration in Fig.~\ref{RamanLasers1} is given by \cite{Gou2020prl,Zhou2020pra}
\begin{eqnarray}\label{eq:HS0}
\hat{H}_{S} = \frac{\hbar^{2}\hat{k}^{2}}{2m}(|A\rangle\langle A| + |B\rangle\langle B|)
+\hbar\omega_{A}|A\rangle\langle A| + \hbar\omega_{B}|B\rangle\langle B| + \hbar\omega_{e}|e\rangle\langle e| + {\bf d}\cdot{\bf E},
\end{eqnarray}
where $|A\rangle$ and $|B\rangle$ are two ground states, $|e\rangle$ is the excited state.
The interaction between the atom and light is the dipole interaction ${\bf d}\cdot{\bf E}$ which is given by
\begin{eqnarray}
{\bf d}\cdot{\bf E}=(d_{Ae}|A\rangle\langle e| + d_{eA}|e\rangle\langle A|)E_{1}(x,t)
+ (d_{Be}|B\rangle\langle e| + d_{eB}|e\rangle\langle B|)E_{2}(x,t), \label{eq:dE}
\end{eqnarray} where $d_{Ae}=-e\langle A|\hat{x}|e\rangle$, $d_{eA}=-e\langle e|\hat{x}|A\rangle$, $d_{Be}=-e\langle B|\hat{x}|e\rangle$, $d_{eB}=-e\langle e|\hat{x}|B\rangle$,
and the electric field ${\bf E}={\bf E}_{1}+{\bf E}_{2}$ is given by
\begin{eqnarray}
E_{1}(x,t)&=&\epsilon_{1}\cos(k_{1}x-\omega_{1}t+\phi_{1})=\frac{\epsilon_{1}}{2}[e^{i(k_{1}x-\omega_{1}t+\phi_{1})} + e^{-i(k_{1}x-\omega_{1}t+\phi_{1})}], \label{eq:E1}\\
E_{2}(x,t)&=&\epsilon_{2}\cos(k_{2}x-\omega_{2}t+\phi_{2})=\frac{\epsilon_{2}}{2}[e^{i(k_{2}x-\omega_{2}t+\phi_{2})} + e^{-i(k_{2}x-\omega_{2}t+\phi_{2})}], \label{eq:E2}
\end{eqnarray} in Fig.~\ref{RamanLasers1}. Notice that the electric field is classical here.

Substituting Eqs.~(\ref{eq:E1}) and (\ref{eq:E2}) into (\ref{eq:dE}), we obtain
\begin{eqnarray}
{\bf d}\cdot{\bf E}&=&\frac{1}{2}\hbar\Omega_{R1}(|A\rangle\langle e| + |e\rangle\langle A|)[e^{i(k_{1}x-\omega_{1}t+\phi_{1})} + e^{-i(k_{1}x-\omega_{1}t+\phi_{1})}] \nonumber\\
&+& \frac{1}{2}\hbar\Omega_{R2}(|B\rangle\langle e| + |e\rangle\langle B| )[e^{i(k_{2}x-\omega_{2}t+\phi_{2})} + e^{-i(k_{2}x-\omega_{2}t+\phi_{2})}],
\end{eqnarray} where the Rabi frequencies are given by the dipole energies
\begin{eqnarray}
\Omega_{R1}&=&\frac{d_{Ae}\epsilon_{1}}{\hbar}=\frac{d_{eA}\epsilon_{1}}{\hbar},\\
\Omega_{R2}&=&\frac{d_{Be}\epsilon_{1}}{\hbar}=\frac{d_{eB}\epsilon_{1}}{\hbar}.
\end{eqnarray}

\begin{figure}[t]
\includegraphics[width=4in]{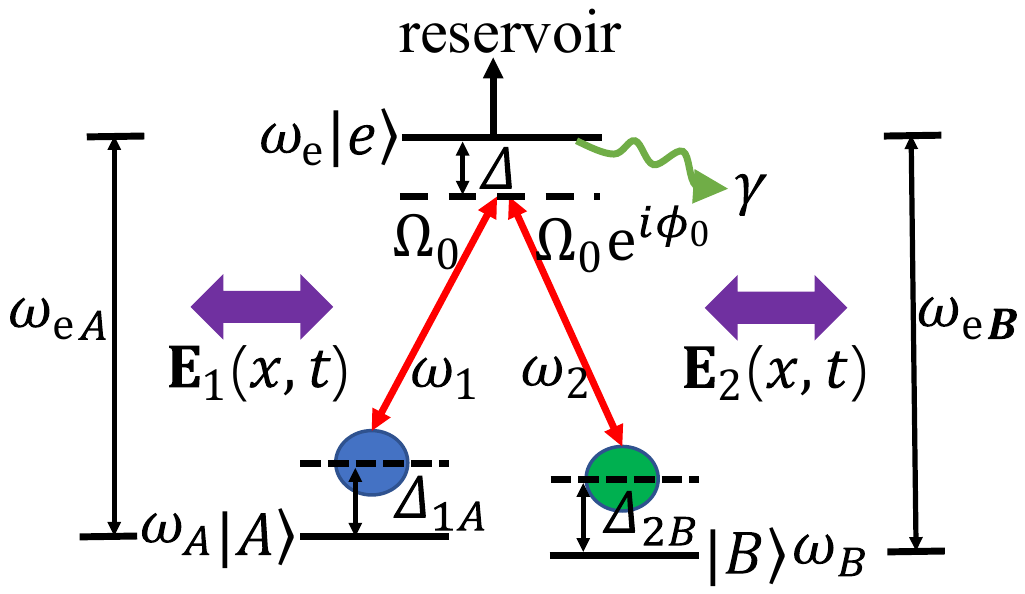}
\caption{Schematic of one unit cell of the experimental setup for the dissipative Raman process, without impurity interaction, with quantities in the derivation indicated. }
\label{RamanLasers1}
\end{figure}

\subsection{Interaction picture}

We next perform the Born-Oppenheimer approximation, where we separate the kinetic energy in Eq.~(\ref{eq:HS0}), and obtain
\begin{eqnarray}\label{eq:HS}
\hat{H}_{S} = \frac{\hbar^{2}\hat{k}^{2}}{2m}(|A\rangle\langle A| + |B\rangle\langle B|) + \hat{H}_{S0} + {\bf d}\cdot{\bf E},
\end{eqnarray} where
\begin{eqnarray}\label{eq:H0}
\hat{H}_{S0}=\hbar\omega_{A}|A\rangle\langle A| + \hbar\omega_{B}|B\rangle\langle B| + \hbar\omega_{e}|e\rangle\langle e|.
\end{eqnarray}

Next we go to the interaction picture, where the Hamiltonian is transformed as
\begin{align}
\hat{H}_{I} &= \hat{U}^{\dagger}(t)\hat{H}_{S}\hat{U}(t) - i\hbar\hat{U}^{\dagger}(t)[\partial_{t}\hat{U}(t)] \\
&=\frac{\hbar^{2}\hat{k}^{2}}{2m}(|A\rangle\langle A| + |B\rangle\langle B|) + \exp\left(\frac{i}{\hbar}\hat{H}_{S0}t\right)({\bf d}\cdot{\bf E})\exp\left(-\frac{i}{\hbar}\hat{H}_{S0}t\right) \nonumber\\
&=\frac{\hbar^{2}\hat{k}^{2}}{2m}(|A\rangle\langle A| + |B\rangle\langle B|) + \frac{1}{2}\hbar\Omega_{R1}\exp\left(\frac{i}{\hbar}\hat{H}_{S0}t\right)(|A\rangle\langle e| + |e\rangle\langle A|)\exp\left(-\frac{i}{\hbar}\hat{H}_{S0}t\right)[e^{i(k_{1}x-\omega_{1}t+\phi_{1})} + e^{-i(k_{1}x-\omega_{1}t+\phi_{1})}] \nonumber\\
&+ \frac{1}{2}\hbar\Omega_{R2}\exp\left(\frac{i}{\hbar}\hat{H}_{S0}t\right)(|B\rangle\langle e| + |e\rangle\langle B|)\exp\left(-\frac{i}{\hbar}\hat{H}_{S0}t\right)[e^{i(k_{2}x-\omega_{2}t+\phi_{2})} + e^{-i(k_{2}x-\omega_{2}t+\phi_{2})}] \nonumber\\
&=\frac{\hbar^{2}\hat{k}^{2}}{2m}(|A\rangle\langle A| + |B\rangle\langle B|) + \frac{1}{2}\hbar\Omega_{R1}\left[e^{i(\omega_{A}-\omega_{e})t}|A\rangle\langle e| + e^{-i(\omega_{A}-\omega_{e})t}|e\rangle\langle A|\right]\left[e^{i(k_{1}x-\omega_{1}t+\phi_{1})} + e^{-i(k_{1}x-\omega_{1}t+\phi_{1})}\right] \nonumber\\
&+ \frac{1}{2}\hbar\Omega_{R2}\left[e^{i(\omega_{B}-\omega_{e})t}|B\rangle\langle e| + e^{-i(\omega_{B}-\omega_{e})t}|e\rangle\langle B|\right]\left[e^{i(k_{2}x-\omega_{2}t+\phi_{2})} + e^{-i(k_{2}x-\omega_{2}t+\phi_{2})}\right] \nonumber\\
&=\frac{\hbar^{2}\hat{k}^{2}}{2m}(|A\rangle\langle A| + |B\rangle\langle B|) + \frac{1}{2}\hbar\Omega_{R1}\left[e^{-i\omega_{eA}t}|A\rangle\langle e| + e^{i\omega_{eA}t}|e\rangle\langle A|\right]\left[e^{i(k_{1}x-\omega_{1}t+\phi_{1})} + e^{-i(k_{1}x-\omega_{1}t+\phi_{1})}\right] \nonumber\\
&+ \frac{1}{2}\hbar\Omega_{R2}\left[e^{-i\omega_{eB}t}|B\rangle\langle e| + e^{i\omega_{eB}t}|e\rangle\langle B|\right]\left[e^{i(k_{2}x-\omega_{2}t+\phi_{2})} + e^{-i(k_{2}x-\omega_{2}t+\phi_{2})}\right] \nonumber\\
&=\frac{\hbar^{2}\hat{k}^{2}}{2m}(|A\rangle\langle A| + |B\rangle\langle B|) + \frac{1}{2}\hbar\Omega_{R1}\left\{\left[e^{i(k_{1}x-(\omega_{1}+\omega_{eA})t+\phi_{1})} + e^{-i(k_{1}x-(\omega_{1}-\omega_{eA})t+\phi_{1})}\right]|A\rangle\langle e| \right. \nonumber\\
&\left.+ \left[e^{i(k_{1}x-(\omega_{1}-\omega_{eA})t+\phi_{1})} + e^{-i(k_{1}x-(\omega_{1}+\omega_{eA})t+\phi_{1})}\right]|e\rangle\langle A|\right\} \nonumber\\
&+ \frac{1}{2}\hbar\Omega_{R2}\left\{\left[e^{i(k_{2}x-(\omega_{2}+\omega_{eB})t+\phi_{2})} + e^{-i(k_{2}x-(\omega_{2}-\omega_{eB})t+\phi_{2})}\right]|B\rangle\langle e| \right. \nonumber\\
&\left.+ \left[e^{i(k_{2}x-(\omega_{2}-\omega_{eB})t+\phi_{2})} + e^{-i(k_{2}x-(\omega_{2}+\omega_{eB})t+\phi_{2})}\right]|e\rangle\langle B|\right\},
\end{align} where we set that $\omega_{eA}=\omega_{e}-\omega_{A}>0$ and $\omega_{eB}=\omega_{e}-\omega_{B}>0$,
\begin{align}
\hat{U}(t)&=\exp\left(-\frac{i}{\hbar}\hat{H}_{S0}t\right),\\
i\hbar\hat{U}^{\dagger}(t)[\partial_{t}\hat{U}(t)]&=i\hbar\exp\left(\frac{i}{\hbar}\hat{H}_{S0}t\right)\left[\partial_{t}\exp\left(-\frac{i}{\hbar}\hat{H}_{S0}t\right)\right]=\hat{H}_{S0}.
\end{align}

\subsection{Rotating-wave approximation}

We next get rid of very rapidly oscillating phases by performing the rotating-wave approximation. Ignoring the counter rotating terms proportional to $e^{\pm i(\omega_{1/2}+\omega_{A/B})t}$, we obtain
\begin{align}\label{eq:HI_rotate}
\hat{H}_{I}^{\rm RWA}
&=\frac{\hbar^{2}\hat{k}^{2}}{2m}(|A\rangle\langle A| + |B\rangle\langle B|) + \frac{1}{2}\hbar\Omega_{R1}\left[e^{-i(k_{1}x-\Delta_{1A}t+\phi_{1})}|A\rangle\langle e| + e^{i(k_{1}x-\Delta_{1A}t+\phi_{1})} |e\rangle\langle A|\right] \nonumber\\
&+ \frac{1}{2}\hbar\Omega_{R2}\left[e^{-i(k_{2}x-\Delta_{2B}t+\phi_{2})}|B\rangle\langle e| + e^{i(k_{2}x-\Delta_{2B}t+\phi_{2})} |e\rangle\langle B|\right],
\end{align} where $\Delta_{1A}=\omega_{1}-\omega_{eA}$ and $\Delta_{2B}=\omega_{2}-\omega_{eB}$ are both single-photon detunings.

Then, we transfer the above Hamiltonian back to Schr{\"o}dinger picture:
\begin{align}
\hat{H}_{S}^{\rm RWA}&= \hat{U}(t)\hat{H}_{I}^{\rm RWA}\hat{U}^{\dagger}(t) + i\hbar[\partial_{t}\hat{U}(t)]\hat{U}^{\dagger}(t) \\
&=\hat{U}(t)(\hat{H}_{I}+\hat{H}_{S0})\hat{U}^{\dagger}(t) \nonumber\\
&=\frac{\hbar^{2}\hat{k}^{2}}{2m}(|A\rangle\langle A| + |B\rangle\langle B|) + \hbar\omega_{A}|A\rangle\langle A| + \hbar\omega_{B}|B\rangle\langle B| + \hbar\omega_{e}|e\rangle\langle e| \nonumber\\
& + \frac{1}{2}\hbar\Omega_{R1}\left[e^{-i(k_{1}x-(\Delta_{1A}+\omega_{eA})t+\phi_{1})}|A\rangle\langle e| + e^{i(k_{1}x-(\Delta_{1A}+\omega_{eA})t+\phi_{1})} |e\rangle\langle A|\right] \nonumber\\
&+ \frac{1}{2}\hbar\Omega_{R2}\left[e^{-i(k_{2}x-(\Delta_{2B}+\omega_{eB})t+\phi_{2})}|B\rangle\langle e| + e^{i(k_{2}x-(\Delta_{2B}+\omega_{eB})t+\phi_{2})} |e\rangle\langle B|\right] \nonumber\\
&=\frac{\hbar^{2}\hat{k}^{2}}{2m}(|A\rangle\langle A| + |B\rangle\langle B|) + \hbar\omega_{A}|A\rangle\langle A| + \hbar\omega_{B}|B\rangle\langle B| + \hbar\omega_{e}|e\rangle\langle e| \nonumber\\
& + \frac{1}{2}\hbar\Omega_{R1}\left[e^{-i(k_{1}x-\omega_{1}t+\phi_{1})}|A\rangle\langle e| + e^{i(k_{1}x-\omega_{1}t+\phi_{1})} |e\rangle\langle A|\right] \nonumber\\
&+ \frac{1}{2}\hbar\Omega_{R2}\left[e^{-i(k_{2}x-\omega_{2}t+\phi_{2})}|B\rangle\langle e| + e^{i(k_{2}x-\omega_{2}t+\phi_{2})} |e\rangle\langle B|\right]. \label{eq:HS_rotate}
\end{align}

\subsection{Rotating frame to remove the time dependent terms}

Since our goal is to obtain a time-independent Hamiltonian description, we introduce a time-dependent unitary transformation $\hat{U}_{c}(t)=e^{i\omega_{1}t|A\rangle\langle A|+i\omega_{2}t|B\rangle\langle B|}$, and transform the Hamiltonian (\ref{eq:HS_rotate}) as
\begin{align}
\hat{H} &= \hat{U}_{c}^{\dagger}(t)\hat{H}_{S}^{\rm RWA}\hat{U}_{c}(t) - i\hbar\hat{U}_{c}^{\dagger}(t)[\partial_{t}\hat{U}_{c}(t)] \nonumber\\
&=\frac{\hbar^{2}\hat{k}^{2}}{2m}(|A\rangle\langle A| + |B\rangle\langle B|) + \hbar(\omega_{A}+\omega_{1})|A\rangle\langle A|+\hbar(\omega_{B}+\omega_{2})|B\rangle\langle B| + \hbar\omega_{e}|e\rangle\langle e| \nonumber\\
&+ \frac{1}{2}\hbar\Omega_{R1}\left[e^{-i(k_{1}x+\phi_{1})}|A\rangle\langle e| + e^{i(k_{1}x+\phi_{1})} |e\rangle\langle A|\right]
+ \frac{1}{2}\hbar\Omega_{R2}\left[e^{-i(k_{2}x+\phi_{2})}|B\rangle\langle e| + e^{i(k_{2}x+\phi_{2})} |e\rangle\langle B|\right],\label{h1}
\end{align} where we have used
\begin{align}
\hat{U}_{c}(t)&=\exp\left[i(\omega_{1}|A\rangle\langle A|+\omega_{2}|B\rangle\langle B|)t\right],\\
i\hbar\hat{U}_{c}^{\dagger}(t)[\partial_{t}\hat{U}_{c}(t)]&=i\hbar\exp\left[-i(\omega_{1}|A\rangle\langle A|+\omega_{2}|B\rangle\langle B|)t\right]\left\{\partial_{t}\exp\left[i(\omega_{1}|A\rangle\langle A|+\omega_{2}|B\rangle\langle B|)t\right]\right\}\nonumber\\
&=-\hbar\omega_{1}|A\rangle\langle A|-\hbar\omega_{2}|B\rangle\langle B|.
\end{align}

For simplicity, by defining $\Omega_{0}=\frac{1}{2}\Omega_{R1}=\frac{1}{2}\Omega_{R2}$, and also enforcing $k_{1}=k_{2}$, $\phi_{2}-\phi_{1}=\phi_{0}$, and applying the gauge transformation $|e\rangle\rightarrow e^{-i(k_{1}x+\phi_{1})}|e\rangle$ on Eq.~(\ref{h1}), we get
\begin{eqnarray}\label{eq:HI}
\hat{H} &=& \frac{\hbar^{2}\hat{k}^{2}}{2m}(|A\rangle\langle A| + |B\rangle\langle B|) + \hbar(\omega_{A}+\omega_{1})|A\rangle\langle A|+\hbar(\omega_{B}+\omega_{2})|B\rangle\langle B| + \hbar\omega_{e}|e\rangle\langle e| \nonumber\\
~~ &+& \hbar\Omega_{0}|e\rangle\langle A| + \hbar\Omega_{0}|A\rangle\langle e| + \hbar\Omega_{0}e^{i\phi_{0}}|e\rangle\langle B|+ \hbar\Omega_{0}e^{-i\phi_{0}}|B\rangle\langle e|,
\end{eqnarray}
where $\hbar\Omega_{0}$ is the strength of the laser-induced coupling between $|A\rangle$ and $|e\rangle$, and $\hbar\Omega_{0}e^{i\phi_{0}}$ is the strength of the laser-induced coupling between $|B\rangle$ and $|e\rangle$.

\subsection{Lindblad master equation and adiabatic elimination towards an effective Hamiltonian}

Having obtained an effective static Hamiltonian (Eq.~(\ref{eq:HI})), we are now ready to consider the effects of dissipation through the Lindblad master equation formalism. This formalism will allow us to rigorously define the adiabatic elimination of the excited state $|e\rangle$, from which we can extract an effective Hamiltonian in the component basis of $\{|A\rangle,|B\rangle\}$.

Temporarily ignoring the diagonal kinetic term $\frac{\hbar^{2}\hat{k}^{2}}{2m}(|A\rangle\langle A| + |B\rangle\langle B|)$ for notational simplicity, the (Hermitian) single-particle effective Hamiltonian describing the ground states $|A\rangle$ and $|B\rangle$ and the excited state $|e\rangle$ is given by
\begin{eqnarray}
\hat{H}_{r}=\hbar\tilde{\omega}_{A}|A\rangle\langle A| + \hbar\tilde{\omega}_{B}|B\rangle\langle B| + \hbar\omega_{e} |e\rangle\langle e|+\left(\hbar\Omega_{0}|e\rangle\langle A| + {\rm h.c.} \right) + \left(\hbar\Omega_{0}e^{i\phi_{0}}|e\rangle\langle B| + {\rm h.c.} \right),
\end{eqnarray} where $\tilde{\omega}_{A}=\omega_{A}+\omega_{1}$ and $\tilde{\omega}_{B}=\omega_{B}+\omega_{2}$.
Upon introducing the laser-induced atom loss on the excited state $|e\rangle$~\cite{liang2022dynamic,gou2020tunable} of the system, the dynamics of our setup is given by the following Lindblad master equation involving the Hermitian single-particle Hamiltonian $\hat{H}_{r}$:~\cite{Zhou2020pra,Zhou2021epl,Manzano_2020,pan2021point-gap,kawabata2022entanglement}
\begin{eqnarray}\label{eq:Lindblad_0}
\frac{d\hat{\rho}}{dt} = -i[\hat{H}_{r}, \hat{\rho}] + \hbar\Gamma\left[\hat{S}\hat{\rho}\hat{S}^{\dagger} - \frac{1}{2}\left\{\hat{S}^{\dagger}\hat{S}, \hat{\rho} \right\} \right]
= -i(\hat{H}_{r}\hat{\rho} - \hat{\rho}\hat{H}_{r}) + \hbar\Gamma\left[\hat{S}\hat{\rho}\hat{S}^{\dagger} - \frac{1}{2}\left(\hat{S}^{\dagger}\hat{S}\hat{\rho} + \hat{\rho}\hat{S}^{\dagger}\hat{S} \right) \right]
=\hat{\cal L}\hat{\rho},
\end{eqnarray}
where $\hat{\cal L}$ is defined as the Liouvillian superoperator, $\hat{\rho}=\sum_{m,n}\rho_{mn}|m\rangle\langle n|$ is the density matrix operator~\cite{pan2021point-gap} for the ground states $|A\rangle$ and $|B\rangle$, the excited state $|e\rangle$, and the reservior state $|r\rangle$ with the element $\rho_{mn}=\langle m|\hat{\rho}|n\rangle$,  $[\cdot,\cdot]$ denotes the  commutation operation, $\{\cdot,\cdot\}$ denotes the anticommutation operation, $\hbar\Gamma$ is the overall decay rate of the excited state $|e\rangle$, and $\hat{S}=|r\rangle\langle e|$ is the quantum jump operator with the reservoir state $|r\rangle$.

Explicitly, in the basis $\left(\langle A|, \langle B|, \langle r|, \langle e| \right)^{T}$, the time evolution equation of each element of the density matrix operator is given by~\cite{Zhou2021epl}
\begin{eqnarray}\label{eq:Lindblad_ij}
\frac{d\rho_{mn}}{dt}
= -i(\hat{\cal H}_{r}\hat{\rho} - \hat{\rho}\hat{\cal H}_{r})_{mn} + \hbar\Gamma\left[\hat{\cal S}\hat{\rho}\hat{\cal S}^{\dagger} - \frac{1}{2}\left(\hat{\cal S}^{\dagger}\hat{\cal S}\hat{\rho} + \hat{\rho}\hat{\cal S}^{\dagger}\hat{\cal S} \right) \right]_{mn},
\end{eqnarray} i.e.,
\begin{eqnarray}
\frac{d\rho_{AA}}{dt}&=&-i\hbar\Omega_{0}(\rho_{eA}-\rho_{Ae}),\label{eq:rho_AA} \\
\frac{d\rho_{AB}}{dt}&=&-i\hbar[(\tilde{\omega}_{A}-\tilde{\omega}_{B})\rho_{AB}+\Omega_{0}(\rho_{eB}-e^{i\phi_0}\rho_{Ae})],\label{eq:rho_AB} \\
\frac{d\rho_{BA}}{dt}&=&-i\hbar[(\tilde{\omega}_{B}-\tilde{\omega}_{A})\rho_{BA}+\Omega_{0}(e^{-i\phi_0}\rho_{eA}-\rho_{Be})],\label{eq:rho_BA} \\
\frac{d\rho_{BB}}{dt}&=&-i\hbar\Omega_{0}(e^{-i\phi_0}\rho_{eB}-e^{i\phi_0}\rho_{Be}),\label{eq:rho_BB} \\
\frac{d\rho_{Ar}}{dt}&=&-i\hbar(\tilde{\omega}_{A}\rho_{Ar}+\Omega_{0}\rho_{er}), \\
\frac{d\rho_{rA}}{dt}&=&-i\hbar(-\tilde{\omega}_{A}\rho_{rA}-\Omega_{0}\rho_{re}), \\
\frac{d\rho_{Br}}{dt}&=&-i\hbar(\tilde{\omega}_{B}\rho_{Br}+e^{-i\phi_0}\Omega_{0}\rho_{er}), \\
\frac{d\rho_{rB}}{dt}&=&-i\hbar(-\rho_{rB}\tilde{\omega}_{B}-e^{i\phi_0}\rho_{re}\Omega_{0}),\\
\frac{d\rho_{Ae}}{dt}&=&-i\hbar[(\tilde{\omega}_{A}-\omega_{e})\rho_{Ae}+\Omega_{0}\rho_{ee}-\Omega_{0}\rho_{AA}-e^{-i\phi_0}\Omega_{0}\rho_{AB}]-\frac{\hbar\Gamma}{2}\rho_{Ae}, \\
\frac{d\rho_{eA}}{dt}&=&-i\hbar[\Omega_{0}\rho_{AA}+\Omega_{0}e^{i\phi_0}\rho_{BA}+(\omega_{e}-\tilde{\omega}_{A})\rho_{eA}-\Omega_{0}\rho_{ee}]-\frac{\hbar\Gamma}{2}\rho_{eA}, \\
\frac{d\rho_{Be}}{dt}&=&-i\hbar[(\tilde{\omega}_{B}-\omega_{e})\rho_{Be}+e^{-i\phi_0}\Omega_{0}\rho_{ee}-\Omega_{0}\rho_{BA}-e^{-i\phi_0}\Omega_{0}\rho_{BB}]-\frac{\hbar\Gamma}{2}\rho_{Be}, \\
\frac{d\rho_{eB}}{dt}&=&-i\hbar(\Omega_{0}\rho_{AB}+\Omega_{0}e^{i\phi_0}\rho_{BB}+(\omega_{e}-\tilde{\omega}_{B})\rho_{eB}-e^{i\phi_0}\rho_{ee}\Omega_{0})-\frac{\hbar\Gamma}{2}\rho_{eB}, \\
\frac{d\rho_{rr}}{dt}&=&\hbar\Gamma\rho_{ee}, \label{eq:rho_rr}\\
\frac{d\rho_{re}}{dt}&=&-i\hbar(-\omega_{e}\rho_{re}-\Omega_{0}\rho_{rA}-e^{-i\phi_0}\Omega_{0}\rho_{rB})-\frac{\hbar\Gamma}{2}\rho_{re}, \\
\frac{d\rho_{er}}{dt}&=&-i\hbar(\Omega_{0}\rho_{Ar}+\Omega_{0}e^{i\phi_0}\rho_{Br}+\omega_{e}\rho_{er})-\frac{\hbar\Gamma}{2}\rho_{er}, \\
\frac{d\rho_{ee}}{dt}&=&-i\hbar\Omega_{0}[(\rho_{Ae}-\rho_{eA}) + (e^{i\phi_0}\rho_{Be}-e^{-i\phi_0}\rho_{eB})]-\hbar\Gamma\rho_{ee},
\end{eqnarray} where we have used
\begin{eqnarray}
\hat{\cal H}_{r}=\begin{pmatrix}
\hbar\tilde{\omega}_{A} & 0 & 0 & \hbar\Omega_{0} \\
0 & \hbar\tilde{\omega}_{B} & 0 & \hbar\Omega_{0}e^{-i\phi_0} \\
0 & 0 & 0 & 0 \\
\hbar\Omega_{0} & \hbar\Omega_{0}e^{i\phi_0} & 0 & \hbar\omega_{e}
\end{pmatrix},~~
\hat{\rho}=\begin{pmatrix}
\rho_{AA} & \rho_{AB} & \rho_{Ar} & \rho_{Ae} \\
\rho_{BA} & \rho_{BB} & \rho_{Br} & \rho_{Be} \\
\rho_{rA} & \rho_{rB} & \rho_{rr} & \rho_{re} \\
\rho_{eA} & \rho_{eB} & \rho_{er} & \rho_{ee}
\end{pmatrix},~~
\hat{\cal S}=\begin{pmatrix}
0 & 0 & 0 & 0 \\
0 & 0 & 0 & 0 \\
0 & 0 & 0 & 1 \\
0 & 0 & 0 & 0
\end{pmatrix},~~ \hat{\cal S}^{\dagger}=\begin{pmatrix}
0 & 0 & 0 & 0 \\
0 & 0 & 0 & 0 \\
0 & 0 & 0 & 0 \\
0 & 0 & 1 & 0
\end{pmatrix}.
\end{eqnarray}
With these explicit expressions, we can also work out the other terms in Eq.~(\ref{eq:Lindblad_ij}):
\begin{eqnarray}
&&(\hat{\cal H}_{r}\hat{\rho} - \hat{\rho}\hat{\cal H}_{r})/\hbar \nonumber\\
&=&\begin{pmatrix}
\tilde{\omega}_{A}\rho_{AA}+\Omega_{0}\rho_{eA} & \tilde{\omega}_{A}\rho_{AB}+\Omega_{0}\rho_{eB} & \tilde{\omega}_{A}\rho_{Ar}+\Omega_{0}\rho_{er} & \tilde{\omega}_{A}\rho_{Ae}+\Omega_{0}\rho_{ee} \\
\tilde{\omega}_{B}\rho_{BA}+e^{-i\phi_0}\Omega_{0}\rho_{eA} & \tilde{\omega}_{B}\rho_{BB}+e^{-i\phi_0}\Omega_{0}\rho_{eB} & \tilde{\omega}_{B}\rho_{Br}+e^{-i\phi_0}\Omega_{0}\rho_{er} & \tilde{\omega}_{B}\rho_{Be}+e^{-i\phi_0}\Omega_{0}\rho_{ee} \\
0 & 0 & 0 & 0 \\
\Omega_{0}\rho_{AA}+\Omega_{0}e^{i\phi_0}\rho_{BA}+\omega_{e}\rho_{eA} & \Omega_{0}\rho_{AB}+\Omega_{0}e^{i\phi_0}\rho_{BB}+\omega_{e}\rho_{eB} & \Omega_{0}\rho_{Ar}+\Omega_{0}e^{i\phi_0}\rho_{Br}+\omega_{e}\rho_{er} & \Omega_{0}\rho_{Ae}+\Omega_{0}e^{i\phi_0}\rho_{Be}+\omega_{e}\rho_{ee}
\end{pmatrix} \nonumber\\
&-&\begin{pmatrix}
\rho_{AA}\tilde{\omega}_{A}+\rho_{Ae}\Omega_{0} & \rho_{AB}\tilde{\omega}_{B}+e^{i\phi_0}\rho_{Ae}\Omega_{0} & 0 & \rho_{AA}\Omega_{0}+e^{-i\phi_0}\rho_{AB}\Omega_{0}+\rho_{Ae}\omega_{e} \\
\rho_{BA}\tilde{\omega}_{A}+\rho_{Be}\Omega_{0} & \rho_{BB}\tilde{\omega}_{B}+e^{i\phi_0}\rho_{Be}\Omega_{0} & 0 & \rho_{BA}\Omega_{0}+e^{-i\phi_0}\rho_{BB}\Omega_{0}+\rho_{Be}\omega_{e} \\
\rho_{rA}\tilde{\omega}_{A}+\rho_{re}\Omega_{0} & \rho_{rB}\tilde{\omega}_{B}+e^{i\phi_0}\rho_{re}\Omega_{0} & 0 & \rho_{rA}\Omega_{0}+e^{-i\phi_0}\rho_{rB}\Omega_{0}+\rho_{re}\omega_{e} \\
\rho_{eA}\tilde{\omega}_{A}+\rho_{ee}\Omega_{0} & \rho_{eB}\tilde{\omega}_{B}+e^{i\phi_0}\rho_{ee}\Omega_{0} & 0 & \rho_{eA}\Omega_{0}+e^{-i\phi_0}\rho_{eB}\Omega_{0}+\rho_{ee}\omega_{e}
\end{pmatrix} \nonumber\\
&=&\left(\begin{array}{cccc}
\Omega_{0}(\rho_{eA}-\rho_{Ae}) & (\tilde{\omega}_{A}-\tilde{\omega}_{B})\rho_{AB}+\Omega_{0}(\rho_{eB}-e^{i\phi_0}\rho_{Ae}) & \tilde{\omega}_{A}\rho_{Ar}+\Omega_{0}\rho_{er} \\
(\tilde{\omega}_{B}-\tilde{\omega}_{A})\rho_{BA}+\Omega_{0}(e^{-i\phi_0}\rho_{eA}-\rho_{Be}) & \Omega_{0}(e^{-i\phi_0}\rho_{eB}-e^{i\phi_0}\rho_{Be}) & \tilde{\omega}_{B}\rho_{Br}+e^{-i\phi_0}\Omega_{0}\rho_{er} \\
-\tilde{\omega}_{A}\rho_{rA}-\Omega_{0}\rho_{re} & -\rho_{rB}\tilde{\omega}_{B}-e^{i\phi_0}\rho_{re}\Omega_{0} & 0 \\
\Omega_{0}\rho_{AA}+\Omega_{0}e^{i\phi_0}\rho_{BA}+(\omega_{e}-\tilde{\omega}_{A})\rho_{eA}-\Omega_{0}\rho_{ee} & \Omega_{0}\rho_{AB}+\Omega_{0}e^{i\phi_0}\rho_{BB}+(\omega_{e}-\tilde{\omega}_{B})\rho_{eB}-e^{i\phi_0}\rho_{ee}\Omega_{0} & \Omega_{0}\rho_{Ar}+\Omega_{0}e^{i\phi_0}\rho_{Br}+\omega_{e}\rho_{er}
\end{array}\right. \nonumber\\
&&\left.\begin{array}{cc}
(\tilde{\omega}_{A}-\omega_{e})\rho_{Ae}+\Omega_{0}\rho_{ee}-\Omega_{0}\rho_{AA}-e^{-i\phi_0}\Omega_{0}\rho_{AB} \\
(\tilde{\omega}_{B}-\omega_{e})\rho_{Be}+e^{-i\phi_0}\Omega_{0}\rho_{ee}-\Omega_{0}\rho_{BA}-e^{-i\phi_0}\Omega_{0}\rho_{BB} \\
-\omega_{e}\rho_{re}-\Omega_{0}\rho_{rA}-e^{-i\phi_0}\Omega_{0}\rho_{rB} \\
\Omega_{0}(\rho_{Ae}-\rho_{eA}) + \Omega_{0}(e^{i\phi_0}\rho_{Be}-e^{-i\phi_0}\rho_{eB})
\end{array}\right)
\end{eqnarray}
and
\begin{eqnarray}
&&\hat{\cal S}\hat{\rho}\hat{\cal S}^{\dagger} - \frac{1}{2}\left(\hat{\cal S}^{\dagger}\hat{\cal S}\hat{\rho} + \hat{\rho}\hat{\cal S}^{\dagger}\hat{\cal S} \right) \nonumber\\
&=&\begin{pmatrix}
0 & 0 & 0 & 0 \\
0 & 0 & 0 & 0 \\
0 & 0 & 0 & 1 \\
0 & 0 & 0 & 0
\end{pmatrix}
\begin{pmatrix}
\rho_{AA} & \rho_{AB} & \rho_{Ar} & \rho_{Ae} \\
\rho_{BA} & \rho_{BB} & \rho_{Br} & \rho_{Be} \\
\rho_{rA} & \rho_{rB} & \rho_{rr} & \rho_{re} \\
\rho_{eA} & \rho_{eB} & \rho_{er} & \rho_{ee}
\end{pmatrix}
\begin{pmatrix}
0 & 0 & 0 & 0 \\
0 & 0 & 0 & 0 \\
0 & 0 & 0 & 0 \\
0 & 0 & 1 & 0
\end{pmatrix}-\frac{1}{2}\begin{pmatrix}
0 & 0 & 0 & 0 \\
0 & 0 & 0 & 0 \\
0 & 0 & 0 & 0 \\
0 & 0 & 1 & 0
\end{pmatrix}\begin{pmatrix}
0 & 0 & 0 & 0 \\
0 & 0 & 0 & 0 \\
0 & 0 & 0 & 1 \\
0 & 0 & 0 & 0
\end{pmatrix}\begin{pmatrix}
\rho_{AA} & \rho_{AB} & \rho_{Ar} & \rho_{Ae} \\
\rho_{BA} & \rho_{BB} & \rho_{Br} & \rho_{Be} \\
\rho_{rA} & \rho_{rB} & \rho_{rr} & \rho_{re} \\
\rho_{eA} & \rho_{eB} & \rho_{er} & \rho_{ee}
\end{pmatrix} \nonumber\\
&-&\frac{1}{2}\begin{pmatrix}
\rho_{AA} & \rho_{AB} & \rho_{Ar} & \rho_{Ae} \\
\rho_{BA} & \rho_{BB} & \rho_{Br} & \rho_{Be} \\
\rho_{rA} & \rho_{rB} & \rho_{rr} & \rho_{re} \\
\rho_{eA} & \rho_{eB} & \rho_{er} & \rho_{ee}
\end{pmatrix}\begin{pmatrix}
0 & 0 & 0 & 0 \\
0 & 0 & 0 & 0 \\
0 & 0 & 0 & 0 \\
0 & 0 & 1 & 0
\end{pmatrix}\begin{pmatrix}
0 & 0 & 0 & 0 \\
0 & 0 & 0 & 0 \\
0 & 0 & 0 & 1 \\
0 & 0 & 0 & 0
\end{pmatrix} \nonumber\\
&=&\begin{pmatrix}
0 & 0 & 0 & 0 \\
0 & 0 & 0 & 0 \\
0 & 0 & 0 & 1 \\
0 & 0 & 0 & 0
\end{pmatrix}\begin{pmatrix}
0 & 0 & \rho_{Ae} & 0 \\
0 & 0 & \rho_{Be} & 0 \\
0 & 0 & \rho_{re} & 0 \\
0 & 0 & \rho_{ee} & 0
\end{pmatrix}-\frac{1}{2}\begin{pmatrix}
0 & 0 & 0 & 0 \\
0 & 0 & 0 & 0 \\
0 & 0 & 0 & 0 \\
0 & 0 & 0 & 1
\end{pmatrix}\begin{pmatrix}
\rho_{AA} & \rho_{AB} & \rho_{Ar} & \rho_{Ae} \\
\rho_{BA} & \rho_{BB} & \rho_{Br} & \rho_{Be} \\
\rho_{rA} & \rho_{rB} & \rho_{rr} & \rho_{re} \\
\rho_{eA} & \rho_{eB} & \rho_{er} & \rho_{ee}
\end{pmatrix} -\frac{1}{2}\begin{pmatrix}
\rho_{AA} & \rho_{AB} & \rho_{Ar} & \rho_{Ae} \\
\rho_{BA} & \rho_{BB} & \rho_{Br} & \rho_{Be} \\
\rho_{rA} & \rho_{rB} & \rho_{rr} & \rho_{re} \\
\rho_{eA} & \rho_{eB} & \rho_{er} & \rho_{ee}
\end{pmatrix}\begin{pmatrix}
0 & 0 & 0 & 0 \\
0 & 0 & 0 & 0 \\
0 & 0 & 0 & 0 \\
0 & 0 & 0 & 1
\end{pmatrix} \nonumber\\
&=& \begin{pmatrix}
0 & 0 & 0 & 0 \\
0 & 0 & 0 & 0 \\
0 & 0 & \rho_{ee} & 0 \\
0 & 0 & 0 & 0
\end{pmatrix}-\frac{1}{2}\begin{pmatrix}
0 & 0 & 0 & 0 \\
0 & 0 & 0 & 0 \\
0 & 0 & 0 & 0 \\
\rho_{eA} & \rho_{eB} & \rho_{er} & \rho_{ee}
\end{pmatrix}-\frac{1}{2}\begin{pmatrix}
0 & 0 & 0 & \rho_{Ae} \\
0 & 0 & 0 & \rho_{Be} \\
0 & 0 & 0 & \rho_{re} \\
0 & 0 & 0 & \rho_{ee}
\end{pmatrix} \nonumber\\
&=& \begin{pmatrix}
0 & 0 & 0 & 0 \\
0 & 0 & 0 & 0 \\
0 & 0 & \rho_{ee} & 0 \\
0 & 0 & 0 & 0
\end{pmatrix}-\frac{1}{2}\begin{pmatrix}
0 & 0 & 0 & \rho_{Ae} \\
0 & 0 & 0 & \rho_{Be} \\
0 & 0 & 0 & \rho_{re} \\
\rho_{eA} & \rho_{eB} & \rho_{er} & 2\rho_{ee}
\end{pmatrix}.
\end{eqnarray}

To obtain the non-Hermitian effective single-particle Hamiltonian $\hat{\mathcal{H}}_R$ in the subspace of the two ground states $\left(\langle A|, \langle B| \right)^{T}$, we perform adiabatic elimination to eliminate the excited state $|e\rangle$ adiabatically in the limit of large detuning $\sqrt{\Delta^{2}+(\Gamma/2)^{2}}\gg\Omega_{0}$.
Physically, adiabatic elimination is used to produce an effective Hamiltonian for a relevant subspace of states, which incorporates effects of its coupling with other states of much higher unperturbed energy. Here,
the adiabatical elimination of the excited state $|e\rangle$ is achieved by setting the following time derivatives to zero, i.e., that all components of the density matrix $\hat{\rho}$ that couple to $|e\rangle$ are constant:
\begin{eqnarray}
\frac{d\rho_{Ae}}{dt}&=&-i\hbar[(\tilde{\omega}_{A}-\omega_{e})\rho_{Ae}+\Omega_{0}\rho_{ee}-\Omega_{0}\rho_{AA}-e^{-i\phi_0}\Omega_{0}\rho_{AB}]-\frac{\hbar\Gamma}{2}\rho_{Ae}=0, \\
\frac{d\rho_{eA}}{dt}&=&-i\hbar[\Omega_{0}\rho_{AA}+\Omega_{0}e^{i\phi_0}\rho_{BA}+(\omega_{e}-\tilde{\omega}_{A})\rho_{eA}-\Omega_{0}\rho_{ee}]-\frac{\hbar\Gamma}{2}\rho_{eA}=0, \\
\frac{d\rho_{Be}}{dt}&=&-i\hbar[(\tilde{\omega}_{B}-\omega_{e})\rho_{Be}+e^{-i\phi_0}\Omega_{0}\rho_{ee}-\Omega_{0}\rho_{BA}-e^{-i\phi_0}\Omega_{0}\rho_{BB}]-\frac{\hbar\Gamma}{2}\rho_{Be}=0, \\
\frac{d\rho_{eB}}{dt}&=&-i\hbar[\Omega_{0}\rho_{AB}+\Omega_{0}e^{i\phi_0}\rho_{BB}+(\omega_{e}-\tilde{\omega}_{B})\rho_{eB}-e^{i\phi_0}\rho_{ee}\Omega_{0}]-\frac{\hbar\Gamma}{2}\rho_{eB}=0, \\
\frac{d\rho_{re}}{dt}&=&-i\hbar(-\omega_{e}\rho_{re}-\Omega_{0}\rho_{rA}-e^{-i\phi_0}\Omega_{0}\rho_{rB})-\frac{\hbar\Gamma}{2}\rho_{re}=0, \\
\frac{d\rho_{er}}{dt}&=&-i\hbar(\Omega_{0}\rho_{Ar}+\Omega_{0}e^{i\phi_0}\rho_{Br}+\omega_{e}\rho_{er})-\frac{\hbar\Gamma}{2}\rho_{er}=0, \\
\frac{d\rho_{ee}}{dt}&=&-i\hbar\Omega_{0}[(\rho_{Ae}-\rho_{eA}) + (e^{i\phi_0}\rho_{Be}-e^{-i\phi_0}\rho_{eB})]-\hbar\Gamma\rho_{ee}=0.\label{eq:Lindblad_ij_eliminate_ee_1}
\end{eqnarray}

\noindent These constraints above can be expressed as the following seven algebraic equations
\begin{eqnarray}
\!\left[\frac{\Gamma}{2}\!-\!i(\omega_{e}\!-\!\tilde{\omega}_{A})\right]\!\rho_{Ae}\!&=&\!i\Omega_{0}(\rho_{AA}\!+\!e^{-i\phi_0}\rho_{AB}\!-\!\rho_{ee}),~\rho_{Ae}\!=\!\frac{-\Omega_{0}(\rho_{AA}\!+\!e^{-i\phi_0}\rho_{AB}\!-\!\rho_{ee})}{(\omega_{e}-\tilde{\omega}_{A})+i\frac{\Gamma}{2}}\!=\!\frac{-\Omega_{0}(\rho_{AA}\!+\!e^{-i\phi_0}\rho_{AB}\!-\!\rho_{ee})}{\Delta_{a}^{\ast}}, \label{eq:Lindblad_ij_eliminate_Ae}\\
\!\left[\frac{\Gamma}{2}\!+\!i(\omega_{e}\!-\!\tilde{\omega}_{A})\right]\!\rho_{eA}\!&=&\!-i\Omega_{0}(\rho_{AA}\!+\!e^{i\phi_0}\rho_{BA}\!-\!\rho_{ee}),~\rho_{eA}\!=\!\frac{-\Omega_{0}(\rho_{AA}\!+\!e^{i\phi_0}\rho_{BA}\!-\!\rho_{ee})}{(\omega_{e}-\tilde{\omega}_{A})-i\frac{\Gamma}{2}}\!=\!\frac{-\Omega_{0}(\rho_{AA}\!+\!e^{i\phi_0}\rho_{BA}\!-\!\rho_{ee})}{\Delta_{a}}, \label{eq:Lindblad_ij_eliminate_eA}\\
\!\left[\frac{\Gamma}{2}\!-\!i(\omega_{e}\!-\!\tilde{\omega}_{B})\right]\!\rho_{Be}\!&=&\!i\Omega_{0}(\tilde{\rho}_{BA}\!+\!e^{-i\phi_0}\rho_{BB}\!-\!e^{-i\phi_0}\rho_{ee}),\nonumber\\
\rho_{Be}\!&=&\!\frac{\!-\!\Omega_{0}(\rho_{BA}\!+\!e^{-i\phi_0}\rho_{BB}\!-\!e^{-i\phi_0}\rho_{ee})}{(\omega_{e}-\tilde{\omega}_{B})+i\frac{\Gamma}{2}}\!=\!\frac{\!-\!\Omega_{0}(\rho_{BA}\!+\!e^{-i\phi_0}\rho_{BB}\!-\!e^{-i\phi_0}\rho_{ee})}{\Delta_{a}^{\ast}}\!, \label{eq:Lindblad_ij_eliminate_Be}\\
\!\left[\frac{\Gamma}{2}\!+\!i(\omega_{e}\!-\!\tilde{\omega}_{B})\right]\!\rho_{eB}\!&=&\!-i\Omega_{0}(\rho_{AB}\!+\!e^{i\phi_0}\rho_{BB}\!-\!e^{i\phi_0}\rho_{ee}),\nonumber\\
\rho_{eB}\!&=&\!\frac{-\Omega_{0}(\rho_{AB}\!+\!e^{i\phi_0}\rho_{BB}\!-\!e^{i\phi_0}\rho_{ee})}{(\omega_{e}-\tilde{\omega}_{B})-i\frac{\Gamma}{2}}\!=\!\frac{-\Omega_{0}(\rho_{AB}\!+\!e^{i\phi_0}\rho_{BB}\!-\!e^{i\phi_0}\rho_{ee})}{\Delta_{a}}, \label{eq:Lindblad_ij_eliminate_eB} \\
\left(\frac{\Gamma}{2}-i\omega_{e}\right)\rho_{re}&=&i\Omega_{0}(\rho_{rA}+e^{-i\phi_0}\rho_{rB}),~~\rho_{re}=\frac{-\Omega_{0}(\rho_{rA}+e^{-i\phi_0}\rho_{rB})}{\omega_{e}+i\frac{\Gamma}{2}}, \label{eq:Lindblad_ij_eliminate_re}\\
\left(\frac{\Gamma}{2}+i\omega_{e}\right)\rho_{er}&=&-i\Omega_{0}(\rho_{Ar}+e^{i\phi_0}\rho_{Br}),~~\rho_{er}=\frac{-\Omega_{0}(\rho_{Ar}+e^{i\phi_0}\rho_{Br})}{\omega_{e}-i\frac{\Gamma}{2}}, \label{eq:Lindblad_ij_eliminate_er} \\
\hbar\Gamma\rho_{ee}&=&-i\hbar\Omega_{0}[(\rho_{Ae}-\rho_{eA}) + (e^{i\phi_0}\rho_{Be}-e^{-i\phi_0}\rho_{eB})] \nonumber\\
&=&i\left[\frac{-\hbar\Omega_{0}^{2}(\rho_{AA}+e^{i\phi_0}\rho_{BA}\!-\!\rho_{ee})}{\Delta_{a}} - \frac{-\hbar\Omega_{0}^{2}(\rho_{AA}+e^{-i\phi_0}\rho_{AB}\!-\!\rho_{ee})}{\Delta_{a}^{\ast}} \right.\nonumber\\
&&\left.~~+ \frac{-\hbar\Omega_{0}^{2}(e^{-i\phi_0}\rho_{AB}+\rho_{BB}\!-\!\rho_{ee})}{\Delta_{a}} - \frac{-\hbar\Omega_{0}^{2}(e^{i\phi_0}\rho_{BA}+\rho_{BB}\!-\!\rho_{ee})}{\Delta_{a}^{\ast}} \right] \nonumber\\
&=&i\left[\frac{-\hbar\Omega_{0}^{2}(\rho_{AA}+\rho_{BB}+e^{i\phi_0}\rho_{BA}\!+\!e^{-i\phi_0}\rho_{AB}\!-\!2\rho_{ee})}{\Delta_{a}} - \frac{-\hbar\Omega_{0}^{2}(\rho_{AA}+\rho_{BB}+e^{-i\phi_0}\rho_{AB}\!+\!e^{i\phi_0}\rho_{BA}\!-\!2\rho_{ee})}{\Delta_{a}^{\ast}} \right] \nonumber\\
&=&-\left(\frac{i\hbar\Omega_{0}^{2}}{\Delta_{a}} - \frac{i\hbar\Omega_{0}^{2}}{\Delta_{a}^{\ast}} \right)(\rho_{AA}+\rho_{BB}) - \left(\frac{i\hbar\Omega_{0}^{2}}{\Delta_{a}} - \frac{i\hbar\Omega_{0}^{2}}{\Delta_{a}^{\ast}} \right)(e^{i\phi_0}\rho_{BA}\!+\!e^{-i\phi_0}\rho_{AB}) + 2\left(\frac{i\hbar\Omega_{0}^{2}}{\Delta_{a}} - \frac{i\hbar\Omega_{0}^{2}}{\Delta_{a}^{\ast}} \right)\rho_{ee} \nonumber\\
&=&\hbar\tilde{\Gamma}(\rho_{AA}+\rho_{BB}) + \hbar\tilde{\Gamma}(e^{i\phi_0}\rho_{BA}\!+\!e^{-i\phi_0}\rho_{AB}) - 2\hbar\tilde{\Gamma}\rho_{ee}, \label{eq:Lindblad_ij_eliminate_ee_2}
\end{eqnarray}
where we have set $\tilde{\omega}_{A}=\tilde{\omega}_{B}=\omega_{0}$, $\Delta_{a}=\omega_{e} - \omega_{0} - i\gamma$, $\gamma=\frac{\Gamma}{2}$, and $\tilde{\Gamma}=2{\rm Im}\frac{\Omega_{0}^{2}}{\Delta_{a}}=2{\rm Im}\frac{\Omega_{0}^{2}}{\omega_{e} - \omega_{0} - i\gamma}=2{\rm Im}\frac{\Omega_{0}^{2}}{\Delta - i\gamma}=2{\rm Im}\frac{\Omega_{0}^{2}(\Delta + i\gamma)}{\Delta^{2}+\gamma^{2}}=\frac{2\gamma\Omega_{0}^{2}}{\Delta^{2}+\gamma^{2}}$. In particular, from Eqs.~(\ref{eq:rho_rr}) and (\ref{eq:Lindblad_ij_eliminate_ee_2}), we have
\begin{eqnarray}
\frac{d\rho_{rr}}{dt}=\hbar\Gamma\rho_{ee}=\hbar\tilde{\Gamma}(\rho_{AA}+\rho_{BB}) + \hbar\tilde{\Gamma}(e^{i\phi_0}\rho_{BA}\!+\!e^{-i\phi_0}\rho_{AB}) - 2\hbar\tilde{\Gamma}\rho_{ee}.\label{eq:eq:rho_rr_eliminate}
\end{eqnarray}

By substituting Eqs.~(\ref{eq:Lindblad_ij_eliminate_Ae})-(\ref{eq:Lindblad_ij_eliminate_eB}) into (\ref{eq:rho_AA})-(\ref{eq:rho_BB}), along with Eq.~(\ref{eq:eq:rho_rr_eliminate}), we can obtain the equations of motion for the remaining density matrix elements in the basis $\left(\langle A|, \langle B|, \langle r| \right)^{T}$:
\begin{eqnarray}
\frac{d\rho_{AA}}{dt}&=&-i\hbar\Omega_{0}(\rho_{eA}-\rho_{Ae})
=-i\left[\frac{-\hbar\Omega_{0}^{2}(\rho_{AA}+e^{i\phi_0}\rho_{BA}\!-\!\rho_{ee})}{\Delta_{a}} - \frac{-\hbar\Omega_{0}^{2}(\rho_{AA}+e^{-i\phi_0}\rho_{AB}\!-\!\rho_{ee})}{\Delta_{a}^{\ast}} \right],\label{eq:rho_AA_t} \\
\frac{d\rho_{AB}}{dt}&=&-i\hbar\Omega_{0}(\rho_{eB}-e^{i\phi_0}\rho_{Ae})
=-i\left[\frac{-\hbar\Omega_{0}^{2}(\rho_{AB}+e^{i\phi_0}\rho_{BB}\!-\!e^{i\phi_0}\rho_{ee})}{\Delta_{a}} - \frac{-\hbar\Omega_{0}^{2}(e^{i\phi_0}\rho_{AA}+\rho_{AB}\!-\!e^{i\phi_0}\rho_{ee})}{\Delta_{a}^{\ast}} \right],\label{eq:rho_AB_t} \\
\frac{d\rho_{BA}}{dt}&=&-i\hbar\Omega_{0}(e^{-i\phi_0}\rho_{eA}-\rho_{Be})
=-i\left[\frac{-\hbar\Omega_{0}^{2}(e^{-i\phi_0}\rho_{AA}+\rho_{BA}\!-\!e^{-i\phi_0}\rho_{ee})}{\Delta_{a}} - \frac{-\hbar\Omega_{0}^{2}(\rho_{BA}+e^{-i\phi_0}\rho_{BB}\!-\!e^{-i\phi_0}\rho_{ee})}{\Delta_{a}^{\ast}} \right],\label{eq:rho_BA_t} \\
\frac{d\rho_{BB}}{dt}&=&-i\hbar\Omega_{0}(e^{-i\phi_0}\rho_{eB}-e^{i\phi_0}\rho_{Be})
=-i\left[\frac{-\hbar\Omega_{0}^{2}(e^{-i\phi_0}\rho_{AB}+\rho_{BB}\!-\!\rho_{ee})}{\Delta_{a}} - \frac{-\hbar\Omega_{0}^{2}(e^{i\phi_0}\rho_{BA}+\rho_{BB}\!-\!\rho_{ee})}{\Delta_{a}^{\ast}} \right],\label{eq:rho_BB_t}\\
\frac{d\rho_{rr}}{dt}&=&\hbar\Gamma\rho_{ee}=\hbar\tilde{\Gamma}(\rho_{AA}+\rho_{BB}) + \hbar\tilde{\Gamma}(e^{i\phi_0}\rho_{BA}\!+\!e^{-i\phi_0}\rho_{AB}) - 2\hbar\tilde{\Gamma}\rho_{ee}.\label{eq:rho_rr_t}
\end{eqnarray}
Since we are considering the large detuning limit $\sqrt{\Delta^{2}+(\Gamma/2)^{2}}\gg\Omega_{0}$, the excited state $|e\rangle$ is barely populated, and the population density on the excited state $|e\rangle$ is very small i.e. $\rho_{ee}\ll\rho_{AA}$, $\rho_{ee}\ll\rho_{BB}$, $\rho_{ee}\ll\rho_{AB}$, $\rho_{ee}\ll\rho_{BA}$, so that $\rho_{ee}$ on the right-hand sides of Eqs.~(\ref{eq:rho_AA_t})-(\ref{eq:rho_rr_t}) can be safely neglected~\cite{Zhou2020pra,Zhou2021epl} as follows:
\begin{eqnarray}
\frac{d\rho_{AA}}{dt}&\approx&-i\left[\frac{-\hbar\Omega_{0}^{2}(\rho_{AA}+e^{i\phi_0}\rho_{BA})}{\Delta_{a}} - \frac{-\hbar\Omega_{0}^{2}(\rho_{AA}+e^{-i\phi_0}\rho_{AB})}{\Delta_{a}^{\ast}} \right],\label{eq:rho_AA_t_2} \\
\frac{d\rho_{AB}}{dt}&\approx&-i\left[\frac{-\hbar\Omega_{0}^{2}(\rho_{AB}+e^{i\phi_0}\rho_{BB})}{\Delta_{a}} - \frac{-\hbar\Omega_{0}^{2}(e^{i\phi_0}\rho_{AA}+\rho_{AB})}{\Delta_{a}^{\ast}} \right],\label{eq:rho_AB_t_2} \\
\frac{d\rho_{BA}}{dt}&\approx&-i\left[\frac{-\hbar\Omega_{0}^{2}(e^{-i\phi_0}\rho_{AA}+\rho_{BA})}{\Delta_{a}} - \frac{-\hbar\Omega_{0}^{2}(\rho_{BA}+e^{-i\phi_0}\rho_{BB})}{\Delta_{a}^{\ast}} \right],\label{eq:rho_BA_t_2} \\
\frac{d\rho_{BB}}{dt}&\approx&-i\left[\frac{-\hbar\Omega_{0}^{2}(e^{-i\phi_0}\rho_{AB}+\rho_{BB})}{\Delta_{a}} - \frac{-\hbar\Omega_{0}^{2}(e^{i\phi_0}\rho_{BA}+\rho_{BB})}{\Delta_{a}^{\ast}} \right],\label{eq:rho_BB_t_2}\\
\frac{d\rho_{rr}}{dt}&\approx&\hbar\tilde{\Gamma}(\rho_{AA}+\rho_{BB}),\label{eq:rho_rr_t_2}
\end{eqnarray}
where, for simplicity, we have set $\phi_0=\pi/2$ in Eq.~(\ref{eq:rho_rr_t_2}) to remove the term $\hbar\tilde{\Gamma}(e^{i\phi_0}\rho_{BA}\!+\!e^{-i\phi_0}\rho_{AB})$.

With these simplifications, the system dynamics are described by the Lindblad master equation in a more compact form:
\begin{eqnarray}\label{eq:Lindblad_1}
\frac{d}{dt}\hat{\tilde{\rho}} = -i\left(\hat{H}_{R}\hat{\tilde{\rho}} - \hat{\tilde{\rho}}\hat{H}_{R}^{\dagger} \right) + \hbar\tilde{\Gamma}\sum_{j=A,B}\hat{L}_{j}\hat{\tilde{\rho}}\hat{L}_{j}^{\dagger}=\hat{\cal L}_{0}\hat{\tilde{\rho}},
\end{eqnarray}
where $\hat{\cal L}_{0}$ is defined as the Liouvillian superoperator, $\hat{L}_{A}=|r\rangle\langle A|$ and $\hat{L}_{B}=|r\rangle\langle B|$ are the quantum jump operators, and $\hat{\tilde{\rho}}$ denotes the density matrix operator in the reduced subspace spanned by the states $\left(\langle A|, \langle B|, \langle r| \right)^{T}$. For notational convenience, we explicitly label the matrix elements as follows:
\begin{eqnarray}
\hat{\cal H}_{R}&=&
	\begin{pmatrix}
		H_{AA} & H_{AB} & 0 \\
		H_{BA} & H_{BB} & 0 \\
		0 & 0 & 0
	\end{pmatrix},\label{eq:H_R_0}\\
\hat{\tilde{\rho}}&=&\begin{pmatrix}
		\rho_{AA} & \rho_{AB} & \rho_{Ar} \\
		\rho_{BA} & \rho_{BB} & \rho_{Br} \\
		\rho_{rA} & \rho_{rB} & \rho_{rr}
	\end{pmatrix}.\label{eq:rho_R_0}
\end{eqnarray}
\noindent By substituting Eqs.~(\ref{eq:H_R_0}) and (\ref{eq:rho_R_0}) into the right-hand side of Eq.~(\ref{eq:Lindblad_1}), we can explicitly write down the Lindblad master equation~(\ref{eq:Lindblad_1}) as follows:
\begin{eqnarray}
\hat{\cal H}_{R}\hat{\tilde{\rho}} - \hat{\tilde{\rho}}\hat{\cal H}_{R}^{\dagger} &=&
\begin{pmatrix}
H_{AA} & H_{AB} & 0 \\
H_{BA} & H_{BB} & 0 \\
0 & 0 & 0
\end{pmatrix}\begin{pmatrix}
\rho_{AA} & \rho_{AB} & \rho_{Ar} \\
\rho_{BA} & \rho_{BB} & \rho_{Br} \\
\rho_{rA} & \rho_{rB} & \rho_{rr}
\end{pmatrix} - \begin{pmatrix}
\rho_{AA} & \rho_{AB} & \rho_{Ar} \\
\rho_{BA} & \rho_{BB} & \rho_{Br} \\
\rho_{rA} & \rho_{rB} & \rho_{rr}
\end{pmatrix}\begin{pmatrix}
H_{AA}^{\ast} & H_{BA}^{\ast} & 0 \\
H_{AB}^{\ast} & H_{BB}^{\ast} & 0 \\
0 & 0 & 0
\end{pmatrix} \nonumber\\
&=&
\begin{pmatrix}
H_{AA}\rho_{AA}+H_{AB}\rho_{BA} & H_{AA}\rho_{AB}+H_{AB}\rho_{BB} & H_{AA}\rho_{Ar}+H_{AB}\rho_{Br}  \\
H_{BA}\rho_{AA}+H_{BB}\rho_{BA} & H_{BA}\rho_{AB}+H_{BB}\rho_{BB} & H_{BA}\rho_{Ar}+H_{BB}\rho_{Br}  \\
0 & 0 & 0
\end{pmatrix} \nonumber\\
&-&
\begin{pmatrix}
H_{AA}^{\ast}\rho_{AA}+H_{AB}^{\ast}\rho_{AB} & H_{BA}^{\ast}\rho_{AA}+H_{BB}^{\ast}\rho_{AB} & 0 \\
H_{AA}^{\ast}\rho_{BA}+H_{AB}^{\ast}\tilde{\rho}_{BB} & H_{BA}^{\ast}\rho_{BA}+H_{BB}^{\ast}\rho_{BB} & 0 \\
H_{AA}^{\ast}\rho_{rA}+H_{AB}^{\ast}\tilde{\rho}_{rB} & H_{BA}^{\ast}\rho_{rA}+H_{BB}^{\ast}\rho_{rB} & 0 \\
\end{pmatrix}, \label{eq:HRrho}
\end{eqnarray}
\begin{eqnarray}
\sum_{j=A,B}\hat{L}_{j}\hat{\tilde{\rho}}\hat{L}_{j}^{\dagger}
&=&\begin{pmatrix}
0 & 0 & 0 \\
0 & 0 & 0 \\
1 & 0 & 0
\end{pmatrix}\begin{pmatrix}
\rho_{AA} & \rho_{AB} & \rho_{Ar} \\
\rho_{BA} & \rho_{BB} & \rho_{Br} \\
\rho_{rA} & \rho_{rB} & \rho_{rr}
\end{pmatrix}\begin{pmatrix}
0 & 0 & 1 \\
0 & 0 & 0 \\
0 & 0 & 0
\end{pmatrix} + \begin{pmatrix}
0 & 0 & 0 \\
0 & 0 & 0 \\
0 & 1 & 0
\end{pmatrix}\begin{pmatrix}
\rho_{AA} & \rho_{AB} & \rho_{Ar} \\
\rho_{BA} & \rho_{BB} & \rho_{Br} \\
\rho_{rA} & \rho_{rB} & \rho_{rr}
\end{pmatrix}\begin{pmatrix}
0 & 0 & 0 \\
0 & 0 & 1 \\
0 & 0 & 0
\end{pmatrix} \nonumber\\
&=&\begin{pmatrix}
0 & 0 & 0 \\
0 & 0 & 0 \\
1 & 0 & 0
\end{pmatrix}\begin{pmatrix}
0 & 0 & \rho_{AA} \\
0 & 0 & \rho_{BA} \\
0 & 0 & \rho_{rA}
\end{pmatrix} + \begin{pmatrix}
0 & 0 & 0 \\
0 & 0 & 0 \\
0 & 1 & 0
\end{pmatrix}\begin{pmatrix}
0 & 0 & \rho_{AB} \\
0 & 0 & \rho_{BB} \\
0 & 0 & \rho_{rB}
\end{pmatrix} = \begin{pmatrix}
0 & 0 & 0 \\
0 & 0 & 0 \\
0 & 0 & \rho_{AA}
\end{pmatrix} + \begin{pmatrix}
0 & 0 & 0 \\
0 & 0 & 0 \\
0 & 0 & \rho_{BB}
\end{pmatrix}
= \begin{pmatrix}
0 & 0 & 0 \\
0 & 0 & 0 \\
0 & 0 & \rho_{AA}+\rho_{BB}
\end{pmatrix}.\nonumber\\
\end{eqnarray}
\noindent By substituting Eq.~(\ref{eq:HRrho}) and the equations of motion~(\ref{eq:rho_AA_t_2})-(\ref{eq:rho_rr_t_2}) into the Lindblad master equation (\ref{eq:Lindblad_1}) and comparing, we can recover the matrix elements of $\hat{H}_{R}$:
\begin{eqnarray}
H_{AA}=-\frac{\hbar\Omega_{0}^{2}}{\Delta-i\gamma},~H_{AB}=-\frac{\hbar\Omega_{0}^{2}}{\Delta-i\gamma}e^{i\phi_0},~H_{BA}=-\frac{\hbar\Omega_{0}^{2}}{\Delta-i\gamma}e^{-i\phi_0},~H_{BB}=-\frac{\hbar\Omega_{0}^{2}}{\Delta-i\gamma},
\end{eqnarray}
where we have set that $\Delta=\Delta_a+i\gamma=\omega_{e} - \omega_{0}$, which is the single-photon detuning of the exited state $|e\rangle$. By adding the kinetic energy term $\frac{\hbar^{2}\hat{k}^{2}}{2m}(|A\rangle\langle A| + |B\rangle\langle B|)$ back into the Hamiltonian $\hat{\cal H}_{R}$, the non-Hermitian effective single-particle Hamiltonian $\hat{H}_{\rm eff}$ in the basis $(\langle A|, \langle B|)^{T}$ is given by
\begin{eqnarray}
\hat{H}_{\rm eff} \!=\! \frac{\hbar^{2}\hat{k}^{2}}{2m} \!+\! \hat{\cal H}_{R} \!=\! \left(\!\begin{array}{cc}
\frac{\hbar^{2}k^{2}}{2m} - \frac{\hbar\Omega_{0}^{2}}{\Delta - i\gamma} & -\frac{\hbar\Omega_{0}^{2}}{\Delta - i\gamma}e^{i\phi_{0}} \\
-\frac{\hbar\Omega_{0}^{2}}{\Delta - i\gamma}e^{-i\phi_{0}} &  \frac{\hbar^{2}k^{2}}{2m} - \frac{\hbar\Omega_{0}^{2}}{\Delta - i\gamma}
\end{array}\!\right) \!=\! \left(\!\begin{array}{cc}
\frac{\hbar^{2}k^{2}}{2m} - \frac{\hbar\Delta\Omega_{0}^{2}}{\Delta^{2}+\gamma^{2}} - i\frac{\hbar\gamma\Omega_{0}^{2}}{\Delta^{2}+\gamma^{2}} & - \frac{\hbar\Delta\Omega_{0}^{2}}{\Delta^{2}+\gamma^{2}}e^{i\phi_{0}} - i\frac{\hbar\gamma\Omega_{0}^{2}}{\Delta^{2}+\gamma^{2}}e^{i\phi_{0}} \\
-\frac{\hbar\Delta\Omega_{0}^{2}}{\Delta^{2}+\gamma^{2}}e^{-i\phi_{0}} - i\frac{\hbar\gamma\Omega_{0}^{2}}{\Delta^{2}+\gamma^{2}}e^{-i\phi_{0}} & \frac{\hbar^{2}k^{2}}{2m} - \frac{\hbar\Delta\Omega_{0}^{2}}{\Delta^{2}+\gamma^{2}} - i\frac{\hbar\gamma\Omega_{0}^{2}}{\Delta^{2}+\gamma^{2}}
\end{array}\!\right),
\end{eqnarray}
Further, with the special choice $\phi_{0}=\pi/2$, i.e., $e^{i\phi_{0}}=i$ for simplicity, we can get
\begin{eqnarray}
\hat{H}_{\rm eff} = \left(\begin{array}{cc}
\frac{\hbar^{2}k^{2}}{2m} - \frac{\hbar\Delta\Omega_{0}^{2}}{\Delta^{2}+\gamma^{2}} - i\frac{\hbar\gamma\Omega_{0}^{2}}{\Delta^{2}+\gamma^{2}} & - i\frac{\hbar\Delta\Omega_{0}^{2}}{\Delta^{2}+\gamma^{2}} + \frac{\hbar\gamma\Omega_{0}^{2}}{\Delta^{2}+\gamma^{2}} \\
i\frac{\hbar\Delta\Omega_{0}^{2}}{\Delta^{2}+\gamma^{2}} - \frac{\hbar\gamma\Omega_{0}^{2}}{\Delta^{2}+\gamma^{2}} & \frac{\hbar^{2}k^{2}}{2m} - \frac{\hbar\Delta\Omega_{0}^{2}}{\Delta^{2}+\gamma^{2}} - i\frac{\hbar\gamma\Omega_{0}^{2}}{\Delta^{2}+\gamma^{2}}
\end{array}\right).
\end{eqnarray}
By setting
\begin{eqnarray}
\tilde{\gamma} = - i\frac{\Delta\Omega_{0}^{2}}{\Delta^{2}+\gamma^{2}} + \frac{\gamma\Omega_{0}^{2}}{\Delta^{2}+\gamma^{2}}=\frac{\Omega_0^2}{\gamma+i\Delta},
\end{eqnarray} we arrive at
\begin{eqnarray}\label{eq:Heff-eff}
\hat{H}_{\rm eff} = \left(\begin{array}{cc}
\frac{\hbar^{2}k^{2}}{2m} - i\hbar\tilde{\gamma} & \hbar\tilde{\gamma} \\
-\hbar\tilde{\gamma} & \frac{\hbar^{2}k^{2}}{2m} - i\hbar\tilde{\gamma}
\end{array}\right).
\end{eqnarray}

\subsection{Alternative derivation of non-Hermitian effective single-particle Hamiltonian from a phenomenological model}

In this subsection, we show how the effective single-particle Hamiltonian Eq.~(\ref{eq:Heff-eff}), which we rigorously derived via the Lindblad master equation, can also be obtained by adiabatically eliminating a phenomenological loss term. In the literature~\cite{liang2022dynamic,gou2020tunable}, this phenomenological loss term is often accounted for by introducing laser-induced atom loss on the excited state $|e\rangle$, and the previous subsection can be construed as a justification of it via the Lindblad formalism.

Phenomenologically, we build upon the Hamiltonian (\ref{eq:HI}) by introducing the loss term on the excited state as follows:}
\begin{eqnarray}
\hat{H} &=& \hat{H}_{0} + \hat{H}_{soc}, \nonumber\\
\hat{H}_{0} &=& \left(\frac{\hbar^{2}\hat{k}^{2}}{2m} + \hbar\tilde{\omega}_{A}\right)|A\rangle\langle A| + \left(\frac{\hbar^{2}\hat{k}^{2}}{2m} + \hbar\tilde{\omega}_{B}\right)|B\rangle\langle B| + \hbar(\omega_{e} - i\gamma)|e\rangle\langle e|, \\
\hat{H}_{soc} &=& \left(\hbar\Omega_{0}|e\rangle\langle A| + {\rm h.c.} \right) + \left(\hbar\Omega_{0}e^{i\phi_{0}}|e\rangle\langle B| + {\rm h.c.} \right),
\end{eqnarray} where $\tilde{\omega}_{A}=\omega_{A}+\omega_{1}$, $\tilde{\omega}_{B}=\omega_{B}+\omega_{2}$, and the decay of the excited state $|e\rangle$ is treated phenomenologically by a decay rate $\gamma$~\cite{liang2022dynamic,gou2020tunable}.

Next, we apply second-order perturbation theory to eliminate the excited state $|e\rangle$:~\cite{Mila2010arxiv,Mila2011book,Sakurai1994book}
\begin{eqnarray}
\langle\sigma|\hat{H}|\sigma'\rangle \approx \langle\sigma|\hat{H}_{0}|\sigma'\rangle + \langle\sigma|\hat{H}_{\rm soc}|\sigma'\rangle + \frac{\langle\sigma|\hat{H}_{\rm soc}|e\rangle\langle e|\hat{H}_{\rm soc}|\sigma'\rangle}{E_{0} - \langle e|\hat{H}_{0}|e\rangle}, \label{eq:LD}
\end{eqnarray}
where $E_{0} = \frac{\hbar^{2}k^{2}}{2m} + \hbar\tilde{\omega}_{\sigma}$ is the eigenvalue of $\hat{H}_{0}$ in ground state, i.e., $\hat{H}_{0}|\sigma\rangle=E_{0}|\sigma\rangle$ with $\sigma=A,B$.

After adiabatic elimination of the exited state $|e\rangle$, we obtain the non-Hermitian effective single-particle Hamiltonian in the basis $(\langle A|, \langle B|)^{T}$
\begin{eqnarray}
\hat{H}_{\rm eff} = \left(\begin{array}{cc}
\frac{\hbar^{2}k^{2}}{2m} + \hbar\omega_{0} - \frac{\hbar\Omega_{0}^{2}}{\Delta_{a}} & -\frac{\hbar\Omega_{0}^{2}}{\Delta_{a}}e^{i\phi_{0}} \\
-\frac{\hbar\Omega_{0}^{2}}{\Delta_{a}}e^{-i\phi_{0}} & \frac{\hbar^{2}k^{2}}{2m} + \hbar\omega_{0} - \frac{\hbar\Omega_{0}^{2}}{\Delta_{a}}
\end{array}\right),
\end{eqnarray}
where we set that $\tilde{\omega}_{A}=\tilde{\omega}_{B}=\omega_{0}$ and $\Delta_{a}=\omega_{e} - i\gamma - \omega_{0}$, and assumed that $\Omega_{0}\ll\sqrt{\Delta^{2}+\gamma^{2}}$.

Further, by shifting the zero energy $\hbar\omega_{0}$, we get the non-Hermitian effective single-particle Hamiltonian as
\begin{eqnarray}
\hat{H}_{\rm eff} = \left(\begin{array}{cc}
\frac{\hbar^{2}k^{2}}{2m} - \frac{\hbar\Omega_{0}^{2}}{\Delta - i\gamma} & -\frac{\hbar\Omega_{0}^{2}}{\Delta - i\gamma}e^{i\phi_{0}} \\
-\frac{\hbar\Omega_{0}^{2}}{\Delta - i\gamma}e^{-i\phi_{0}} &  \frac{\hbar^{2}k^{2}}{2m} - \frac{\hbar\Omega_{0}^{2}}{\Delta - i\gamma}
\end{array}\right) = \left(\begin{array}{cc}
\frac{\hbar^{2}k^{2}}{2m} - \frac{\hbar\Delta\Omega_{0}^{2}}{\Delta^{2}+\gamma^{2}} - i\frac{\hbar\gamma\Omega_{0}^{2}}{\Delta^{2}+\gamma^{2}} & - \frac{\hbar\Delta\Omega_{0}^{2}}{\Delta^{2}+\gamma^{2}}e^{i\phi_{0}} - i\frac{\hbar\gamma\Omega_{0}^{2}}{\Delta^{2}+\gamma^{2}}e^{i\phi_{0}} \\
-\frac{\hbar\Delta\Omega_{0}^{2}}{\Delta^{2}+\gamma^{2}}e^{-i\phi_{0}} - i\frac{\hbar\gamma\Omega_{0}^{2}}{\Delta^{2}+\gamma^{2}}e^{-i\phi_{0}} & \frac{\hbar^{2}k^{2}}{2m} - \frac{\hbar\Delta\Omega_{0}^{2}}{\Delta^{2}+\gamma^{2}} - i\frac{\hbar\gamma\Omega_{0}^{2}}{\Delta^{2}+\gamma^{2}}
\end{array}\right),
\end{eqnarray}
where we have set that $\Delta=\omega_{e} - \omega_{0}$, which is the single-photon detuning of the excited state $|e\rangle$.

By setting $\phi_{0}=\pi/2$, i.e., $e^{i\phi_{0}}=i$ as before, we can recover the previously derived single-particle effective Hamiltonian as
\begin{eqnarray}\label{eq:Heff-eff_2}
\hat{H}_{\rm eff} = \left(\begin{array}{cc}
\frac{\hbar^{2}k^{2}}{2m} - \frac{\hbar\Delta\Omega_{0}^{2}}{\Delta^{2}+\gamma^{2}} - i\frac{\hbar\gamma\Omega_{0}^{2}}{\Delta^{2}+\gamma^{2}} & - i\frac{\hbar\Delta\Omega_{0}^{2}}{\Delta^{2}+\gamma^{2}} + \frac{\hbar\gamma\Omega_{0}^{2}}{\Delta^{2}+\gamma^{2}} \\
i\frac{\hbar\Delta\Omega_{0}^{2}}{\Delta^{2}+\gamma^{2}} - \frac{\hbar\gamma\Omega_{0}^{2}}{\Delta^{2}+\gamma^{2}} & \frac{\hbar^{2}k^{2}}{2m} - \frac{\hbar\Delta\Omega_{0}^{2}}{\Delta^{2}+\gamma^{2}} - i\frac{\hbar\gamma\Omega_{0}^{2}}{\Delta^{2}+\gamma^{2}}\end{array}\right)= \left(\begin{array}{cc}
\frac{\hbar^{2}k^{2}}{2m} - i\hbar\tilde{\gamma} & \hbar\tilde{\gamma} \\
-\hbar\tilde{\gamma} & \frac{\hbar^{2}k^{2}}{2m} - i\hbar\tilde{\gamma}
\end{array}\right).
\end{eqnarray} where $\tilde{\gamma} =\frac{\Omega_0^2}{\gamma+i\Delta}$.
By setting
\begin{eqnarray}
\tilde{\gamma} = - i\frac{\Delta\Omega_{0}^{2}}{\Delta^{2}+\gamma^{2}} + \frac{\gamma\Omega_{0}^{2}}{\Delta^{2}+\gamma^{2}}=\frac{\Omega_0^2}{\gamma+i\Delta},
\end{eqnarray} we arrive at
\begin{eqnarray}
\hat{H}_{\rm eff} = \left(\begin{array}{cc}
\frac{\hbar^{2}k^{2}}{2m} - i\hbar\tilde{\gamma} & \hbar\tilde{\gamma} \\
-\hbar\tilde{\gamma} & \frac{\hbar^{2}k^{2}}{2m} - i\hbar\tilde{\gamma}
\end{array}\right).
\end{eqnarray}

\subsection{Tight-binding model}\label{1.1}

To describe $N$ such atoms subject to two-photon Raman processes in a background lattice potential with impurity-atom interaction, we extend
the single-particle effective Hamiltonian to the second-quantized form ($\hbar=1$)
\begin{align}\label{eq:H00}
\hat{H}_{\rm sec}=\sum_{\sigma=A,B}\int dx\hat{\psi}_{\sigma}^{\dagger}(x)\left[ \frac{\hat{p}_{x}^{2}}{2m} - i\tilde{\gamma} + V(x) \right]\hat{\psi}_{\sigma}(x) + \tilde{\gamma}\int dx\left[\hat{\psi}_{A}^{\dagger}(x)\hat{\psi}_{B}(x) - \hat{\psi}_{B}^{\dagger}(x)\hat{\psi}_{A}(x)\right],
\end{align}
where $\hat{\psi}_{\sigma}(x)$ denotes the field operator for annihilating an atom with state $\sigma$ ($\sigma=A,B$) at position $x$, and the background lattice potential is added. This lattice potential is formed by superimposing two standing optical waves with wavelengths $\lambda_2=767$ nm (short lattice) and $\lambda_1=2\lambda_2$ (long lattice) as~\cite{Atala2013np,Folling2007nature}
\begin{align}
V(x)=V_{1}\sin^{2}(k_{1}x+\phi/2) + V_{2}\sin^{2}(k_{2}x+\pi/2),
\end{align} where $k_{1}=2\pi/\lambda_{1}$, $k_{2}=2k_{1}$, $V_{1}$ and $V_{2}$ are the corresponding strengths of the two standing waves. Phase control between the two standing wave fields is realized by the controlling of  $\phi$.

When the Raman term is much smaller than the deep background lattice depth $V_1$ and $V_2$, i.e., $\Delta,|\tilde{\gamma}|\ll V_1,V_2$, the tight-binding approximation is applicable at low temperatures.
When atoms experience ultra-low temperatures, only the lowest band $n=0$ will be populated. Therefore, we expand the field operator $\hat{\psi}_{\sigma}(x)$ in terms of the lowest energy level ($n=0$ or $S$-band) Wannier functions $W_{n=0,\sigma}(x-x_j)$ of the background lattice potential~\cite{Liu2013prl,Liu2014prl,Pan2015prl,Fan2018pra}: $\hat{\psi}_{\sigma}(x)=\sum_{j}\hat{c}_{j,\sigma}W_{n=0,\sigma}(x-x_j)$, where $\hat{c}_{j,\sigma}$ $(\hat{c}_{j,\sigma}^{\dagger})$ is the operator annihilating (creating) an atom in the
lowest band ($n=0$ or $S$ band) at site $j$.
Furthermore, we can write the single-band tight-binding Hamiltonian corresponding to Eq. (\ref{eq:H00}) as
\begin{align}
\hat{H}_{\rm TB}=\sum_{j}\left[ - i\tilde{\gamma}(\hat{c}_{j,A}^{\dagger}\hat{c}_{j,A} + \hat{c}_{j,B}^{\dagger}\hat{c}_{j,B}) + \left(t_{1}+\tilde{\gamma}\right)\hat{c}_{j,A}^{\dagger}\hat{c}_{j,B} + \left(t_{1}-\tilde{\gamma}\right)\hat{c}_{j,B}^{\dagger}\hat{c}_{j,A} + t_{2}(\hat{c}_{j+1,A}^{\dagger}\hat{c}_{j,B} + {\rm H.c.}) \right],
\end{align} where
\begin{align}
t_{1}&=\int dxW_{n=0,A}^{*}(x-x_j)\left[ -\frac{\partial_{x}^{2}}{2m} + V_{1}\sin^{2}(k_{1}x+\phi/2) + V_{2}\sin^{2}(k_{2}x+\pi/2)) \right]W_{n=0,B}(x-x_j),\label{eq:t1}\\
t_{2}&=\int dxW_{n=0,B}^{*}(x-x_j)\left[ -\frac{\partial_{x}^{2}}{2m} + V_{1}\sin^{2}(k_{1}x+\phi/2) + V_{2}\sin^{2}(k_{2}x+\pi/2)) \right]W_{n=0,A}(x-x_{j+1}). \label{eq:t2}
\end{align}
Here $\hat{c}_{j,A/B}$ is the annihilation field operator for the state $|A/B\rangle$, the subscript $j$ denotes the cell, the subscript $A/B$ denotes the site in a cell, $t_1$ is the intra-cell hopping, and $t_2$ is the inter-cell hopping.

\begin{figure}[t]
\includegraphics[width=3.2in]{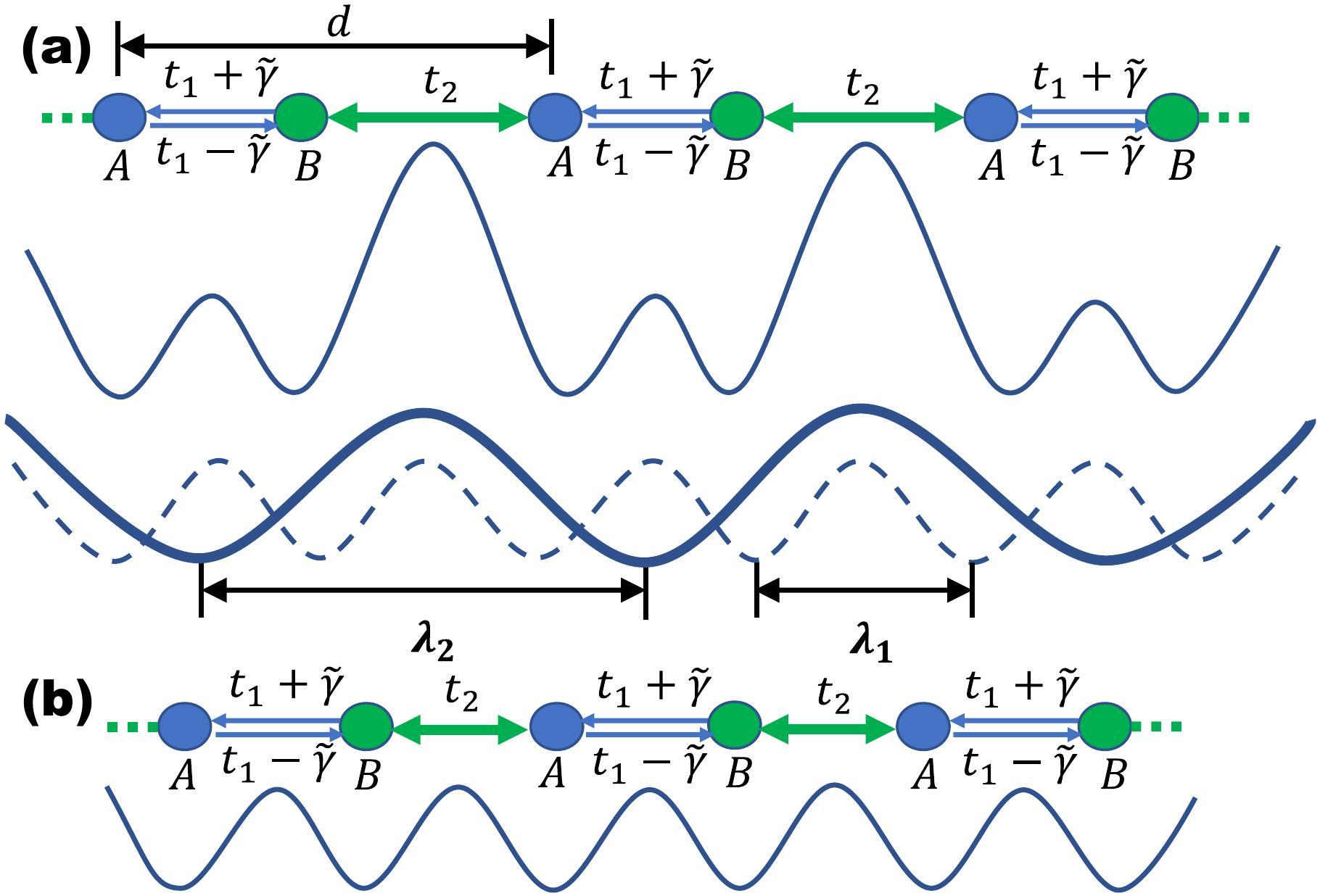}
\caption{(a) Superimposing two optical lattice potentials of different periods creates a 1D array of double-well potentials - shown here is when they differ by a factor of two~\cite{Atala2013np,Folling2007nature}. Their Wannier function overlaps (Eqs.~(\ref{eq:t1}) and (\ref{eq:t2})) give the effective intra- and inter-cell tight-binding hoppings $t_1$ and $t_2$, and the effective decay $\tilde\gamma$ modifies $t_1$ to $t_1\pm\tilde{\gamma}$.
(b) For the special case of $t_1=t_2$, only one optical lattice potential is required.}\label{fig:LatticeS2}
\end{figure}

In our proposed cold atom experiment, the Hamiltonian is in weak-coupling limit ($\Omega_0\ll\sqrt{\Delta^{2}+\gamma^{2}}$, $\Omega_0=-e\langle A|\hat x|e\rangle\epsilon_1/2$) corresponding to the configuration in Fig.~3a1 of the main text. We choose $d=767$ nm~\cite{Atala2013np}, $t_1(t_2)\in(2\pi)\times[60,1000]$ Hz~\cite{Atala2013np}, $\Omega_{0}=(2\pi)\times0.03$ MHz~\cite{Lin2009prl,Lin2011nature,Zhang2012prl,Wang2012prl,Cheuk2012prl,Qu2013pra}. We also set the single-photon detuning of the exited state $|e\rangle$ as $\Delta=(2\pi)\times1$ MHz with the spontaneous decay rate $\gamma=(2\pi)\times6$ MHz~\cite{Fu2013pra,Jie2017pra,Qin2018pra} which create effective decay rate $\tilde{\gamma}\sim(2\pi)\times(0.92-0.15i)$ kHz. We adjust the optical wave length to set dimensional parameters to be $t_{1}/t_{2}=1$ with $t_2=(2\pi)\times 1000$ Hz~\cite{Atala2013np} as the energy unit (Fig.~\ref{fig:LatticeS2}(b)). Then, other parameter takes the value $\tilde{\gamma}/t_{2}=0.92-0.15i$. Performing  Fourier transformation, we get
\begin{align}
\hat{H}_{\rm TB}(k)= - i\tilde{\gamma}{\Gamma}_{2\times2} + (t_{1} + t_{2}\cos k)\sigma_{x} + \left(t_{2}\sin k + i\tilde{\gamma} \right)\sigma_{y},
\end{align} where ${\Gamma}_{2\times2}$ is the $2\times2$ identity matrix.

\subsection{Non-Hermitian interaction model}\label{1.2}

\begin{figure}[t]
\includegraphics[width=2.0in]{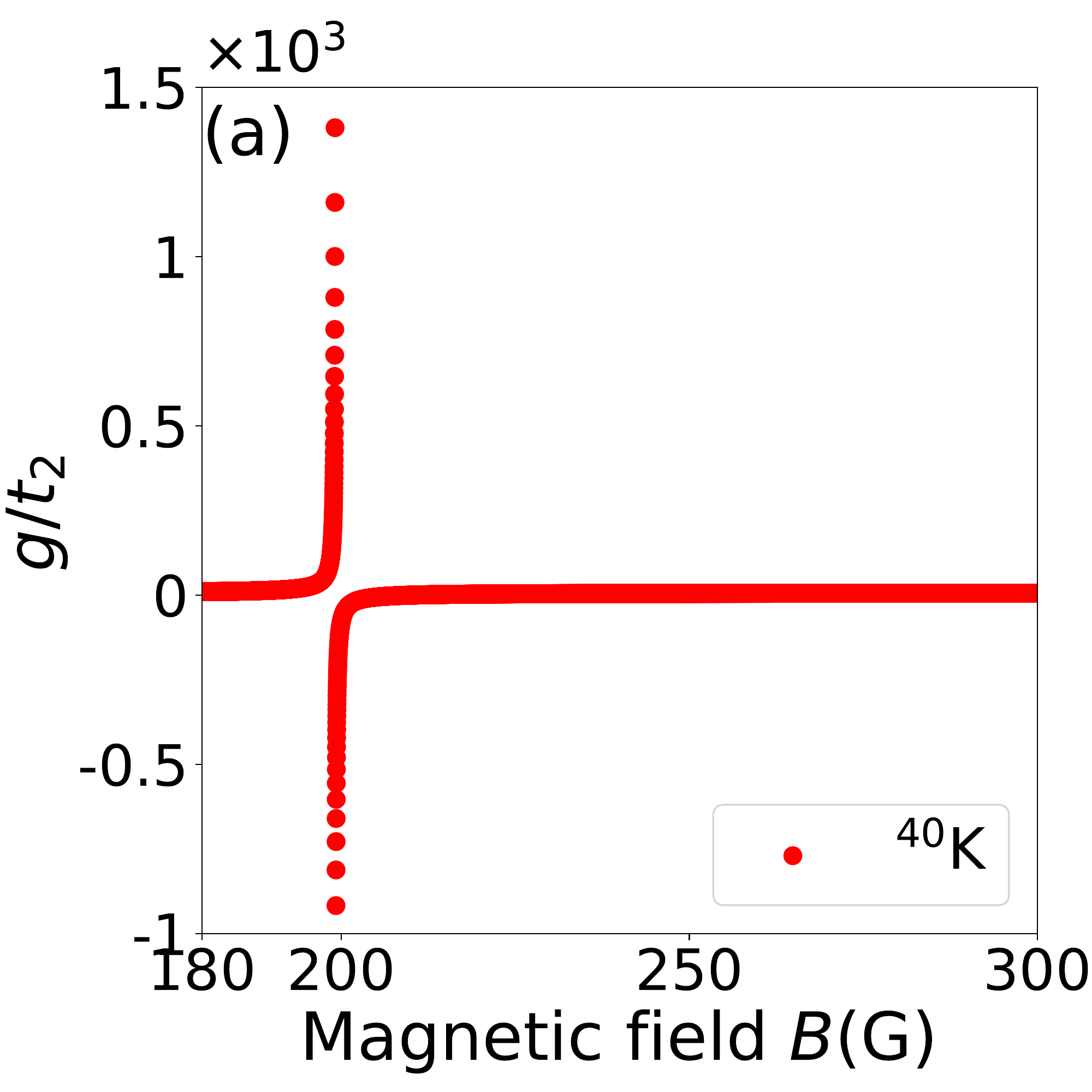}\includegraphics[width=2.0in]{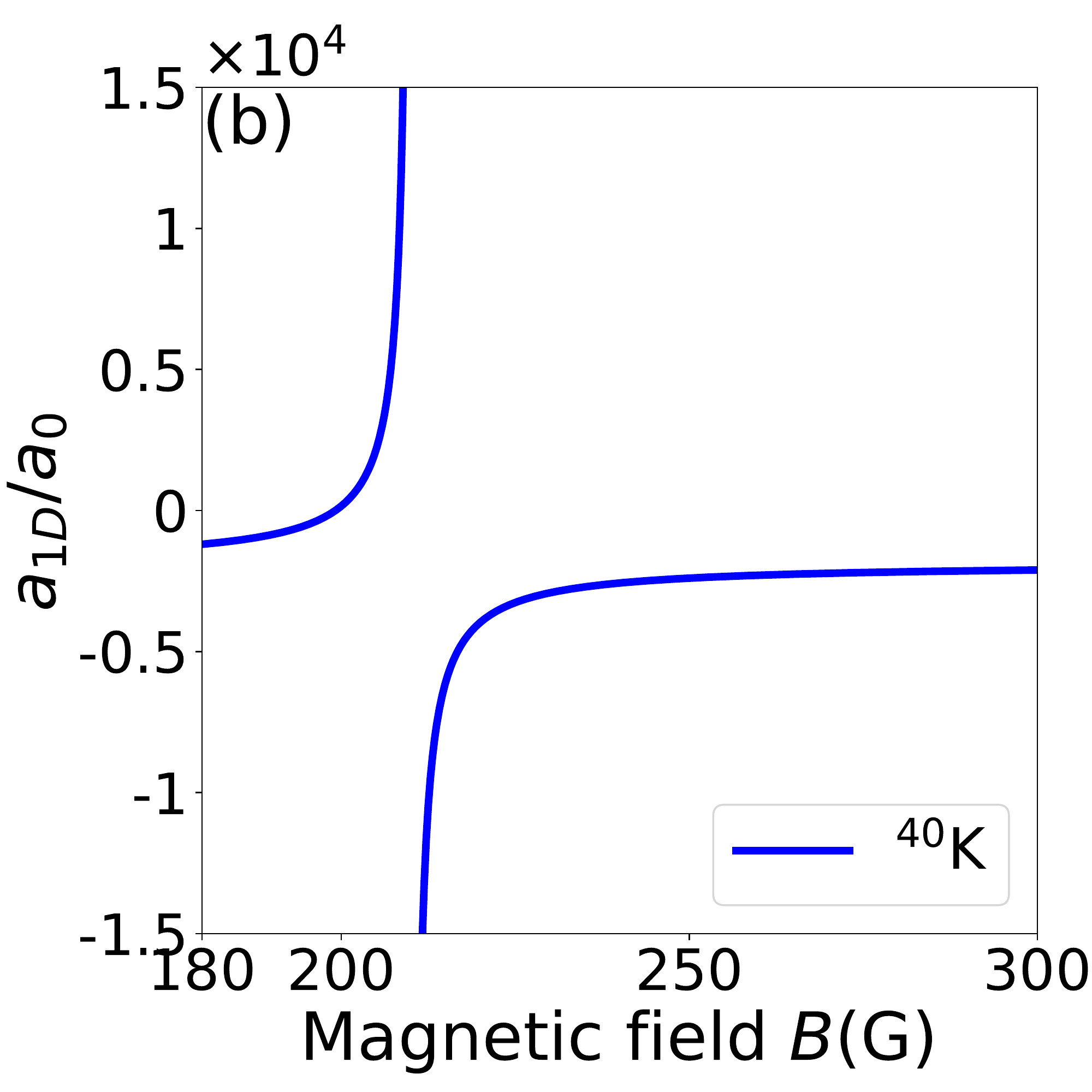}\includegraphics[width=2.0in]{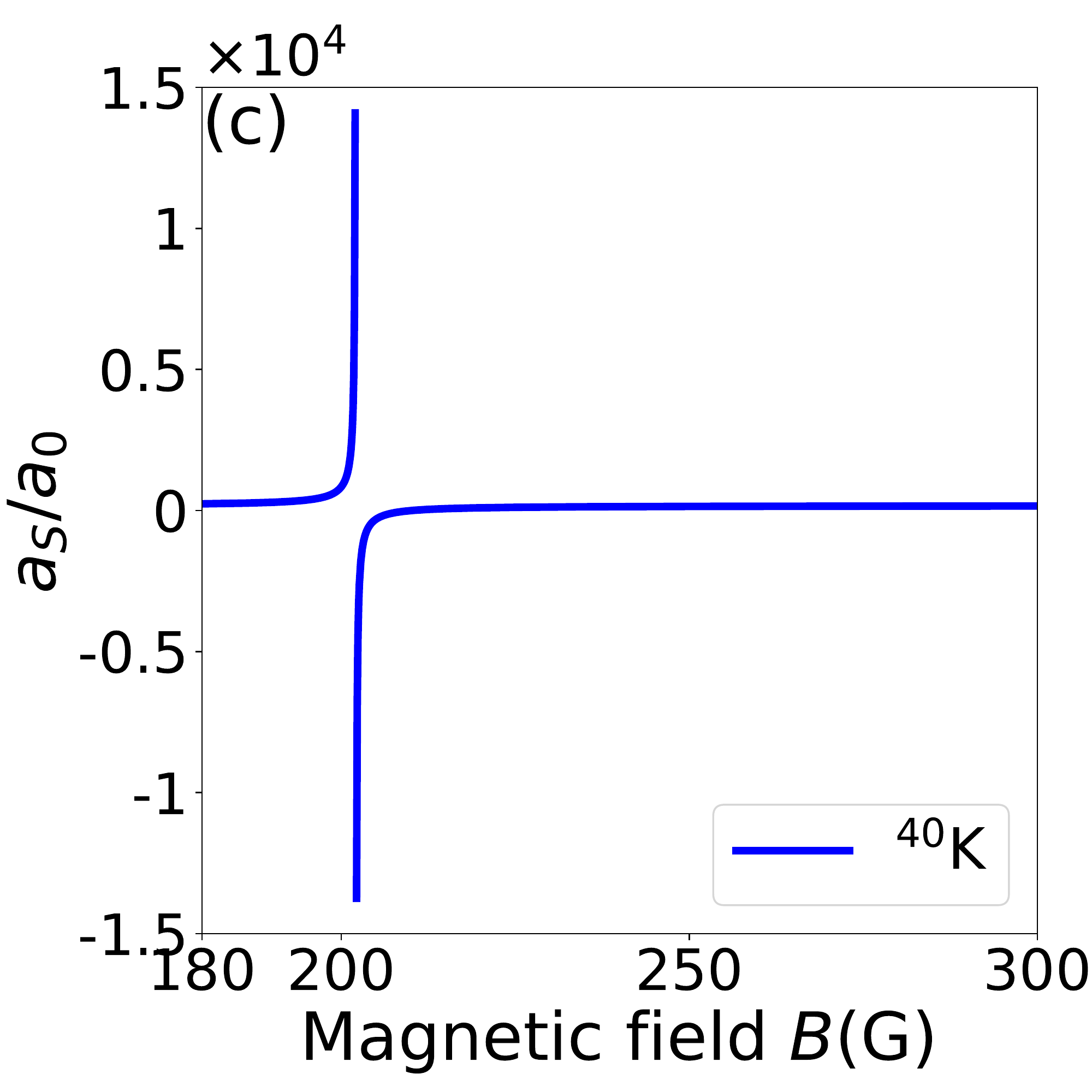}
\caption{(a) Feshbach interaction strength $g$ as a function of the magnetic field for $^{40}$K atoms. (b) Effective one-dimensional $S$-wave scattering length $a_{1D}$ as a function of the magnetic field for $^{40}$K atoms. (c) Three-dimensional $S$-wave scattering length $a_S$ as a function of the magnetic field for $^{40}$K atoms. The experimental parameters are given by $t_2=(2\pi)\times1000$ Hz~\cite{Atala2013np}, $\omega_{\perp}=(2\pi)\times2\times10^5$ Hz~\cite{Liao2010nature} $B_{0}=202.1$G, $\Delta_{B}=8$G, $a_{\rm bg}=174a_{0}$~\cite{Regal2004prl,Chin2010rmp} with the Bohr radius $a_{0}$. }\label{fig:g_K40}
\end{figure}

In order to investigate polaron physics in our setup, we consider Feshbach resonance interaction between majority spin-$\uparrow$ fermions and and a static impurity fermion with spin-$\downarrow$ in $s$ site of the $x_0$-th cell.  Here, we choose $s=A$ in the main text, and there is analogous results if we had chosen $s=B$ instead.

As a result, atoms with spin-$\uparrow$ and spin-$\downarrow$  interact via an on-site interaction Hamiltonian $\hat{H}_{\rm Int}$ in $s$ site of the $x_0$-th cell
\begin{align}
\hat{H}_{\rm int}=g\hat{b}_{x_0,s}^{\dagger}\hat{b}_{x_0,s}\hat{c}_{x_0,s}^{\dagger}\hat{c}_{x_0,s},
\end{align} where $\hat{b}_{x_0,s}$ with $s=A/B$ is the annihilation field operator for the impurity atom in $s$ site of the $x_0$-th cell. The interaction coupling constant for a unit cell of size $d$ takes
\begin{align}\label{eq:g1D}
g=g_{1D}/d,
\end{align}
where
\begin{align}\label{eq:G1D}
g_{1D}=-\frac{\hbar^2}{m_{r}a_{1D}},
\end{align} with the reduced mass $m_{r}=m/2$. Olshanii first obtained the atom-atom scattering under transverse harmonic confinement where the effective one-dimensional coupling strength $g_{1D}$ diverges at a particular ratio of the confinement and scattering lengths $a_S$ as shown in Eqs.~(\ref{eq:a1D}) and (\ref{eq:aS}). Hence the even-wave scattering length $a_{1D}$ (shown in Fig.~\ref{fig:g_K40}(b)) in one dimension is~\cite{Olshanii1998prl,Olshanii2003prl,Chin2010rmp,Qin20201pra,Qin20202pra}
\begin{align}\label{eq:a1D}
a_{1D}=-\frac{\ell^{2}_{\perp}}{2a_S} + \frac{{\cal C}\ell_{\perp}}{2},
\end{align}
where ${\cal C}=1.4603$, $\ell_{\perp}=\sqrt{2\hbar/(m\omega_{\perp})}$, and $\omega_{\perp}$ is the transverse trapping frequency with $\omega_{\perp}=(2\pi)\times2\times10^5$ Hz in experiment~\cite{Liao2010nature}. This gives ${\cal C}\ell_{\perp}/2\approx690.74a_{0}$. As shown in Fig.~\ref{fig:g_K40}(c), near the resonance magnetic field $B_{0}=202.1$G, we have $a_S\gg{\cal C}\ell_{\perp}/2$. $a_S$ is the three-dimensional $S$-wave scattering length which is empirically fitted to be~\cite{Chin2010rmp,Regal2004prl}
\begin{align}\label{eq:aS}
a_{S}=a_{\rm bg}\left(1 - \frac{\Delta_{B}}{B-B_{0}} \right),
\end{align} where $a_{\rm bg}$ is the background scattering length, $B$ is the strength of the magnetic field, $B_{0}$ is the position of the Feshbach resonance, and $\Delta_B$ is the three-dimensional resonance width.
For $^{40}$K, we have $B_{0}=202.1$G, $\Delta_{B}=8$G, $a_{\rm bg}=174a_{0}$ with the Bohr radius $a_{0}$~\cite{Regal2004prl,Chin2010rmp}. As such, as shown in Fig.~\ref{fig:g_K40}(a), when $B\rightarrow B_0$, $g$ does not diverge, but saturates at a large $\sim -17.88t_2$. When $B\rightarrow B_c\sim200$G, $g$ saturates at a minimum $\sim -1000t_2$.

\newpage

\section{Additional data}

\subsection{Energy spectra}\label{4}

\subsubsection{Energy spectra for minimal toy model $\hat H_{\rm min}$ under PBCs}

\begin{figure}[h]
{\includegraphics[width=0.7\linewidth]{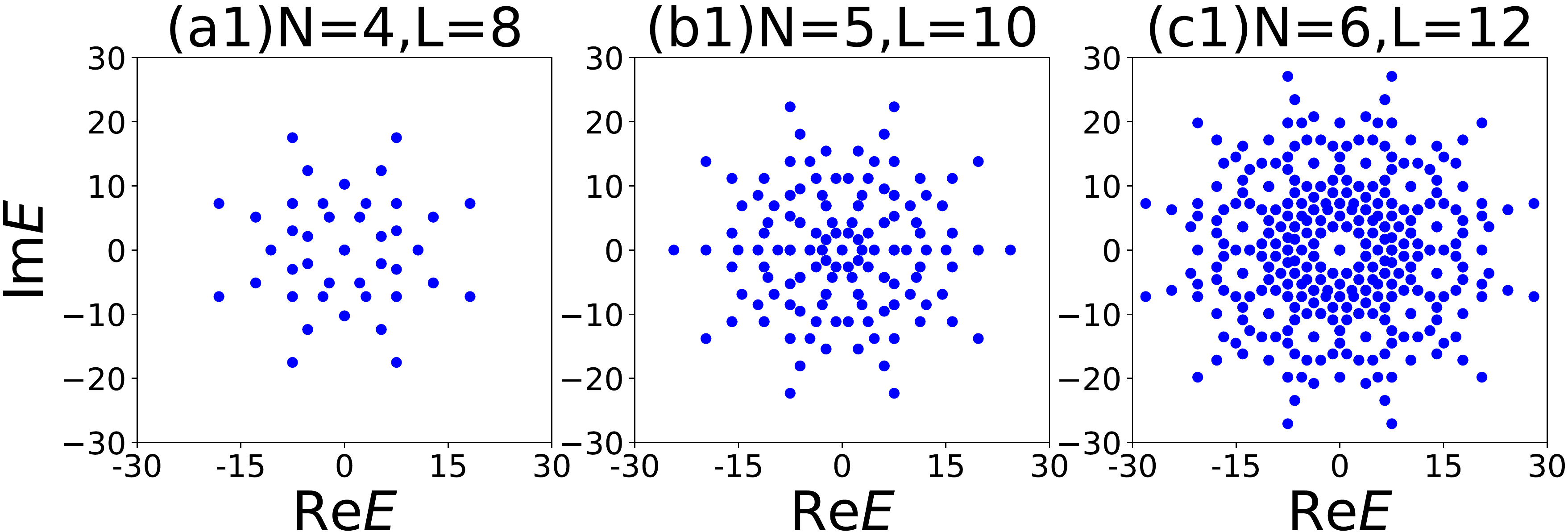}}\\
{\includegraphics[width=0.7\linewidth]{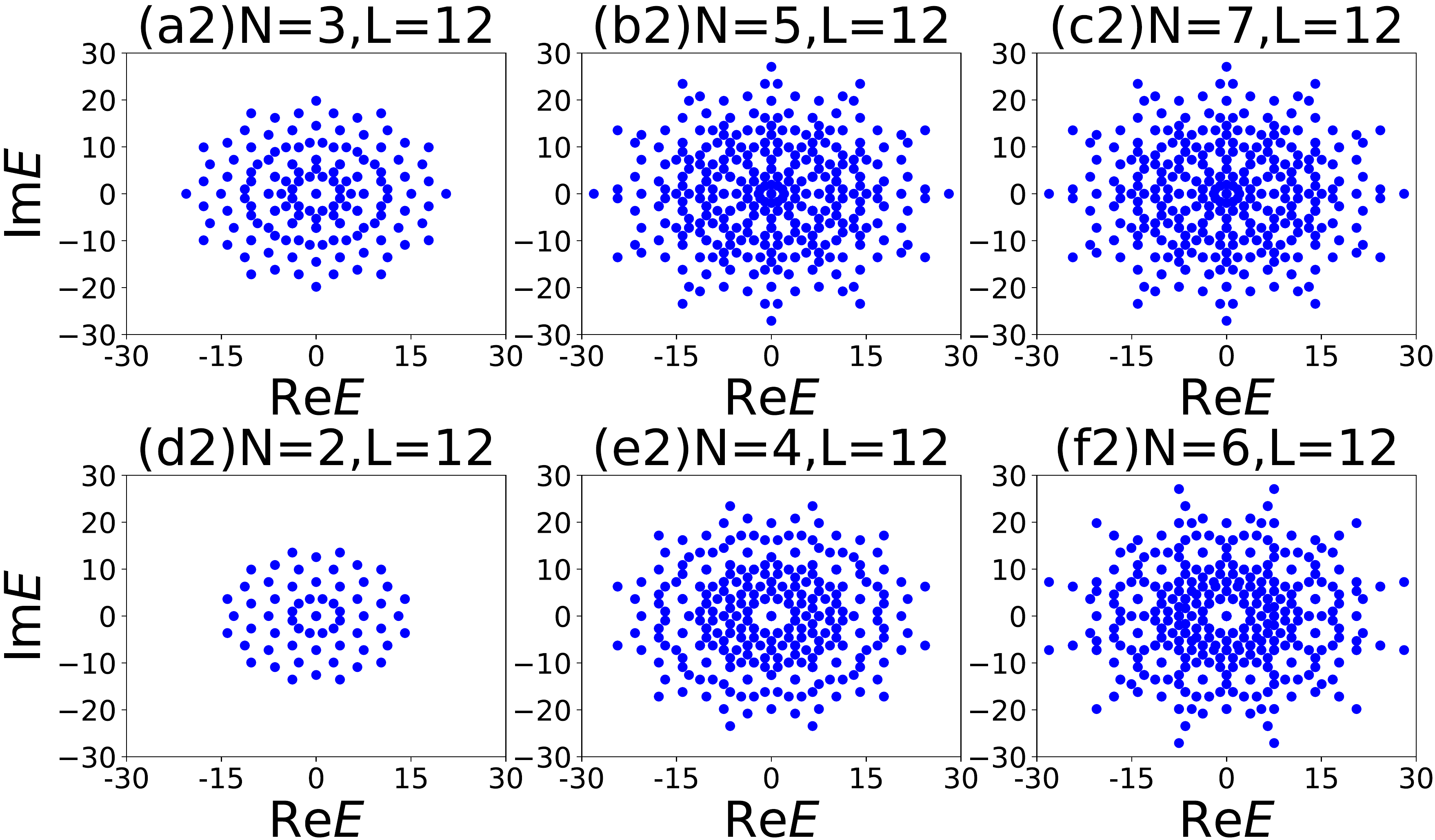}}
\caption{Energy spectra of $\hat H_\text{min}$ under PBCs for different number of fermions $N$ and sites $L$ with $\alpha=2$ and $g=0$. In general, the number of spikes is given by $L$. (a1-c1) Different system sizes $L$, all at half-filling. (a2-f2) Different number of fermions, at constant number of sites $L=12$, where (a2-c2) are for different odd number of fermions and (d2-f2) are for different even number of fermions.}
\label{fig:Hmin_energy_PBC}
\end{figure}

In general, it is indicated from Fig.~\ref{fig:Hmin_energy_PBC} that the number of spikes in the energy spectrum is given by the number of system sizes $L$.

\newpage

\subsubsection{Energy spectra for cold-atom setup $\hat H$ under OBCs}

\begin{figure}[h]
\includegraphics[width=1\linewidth]{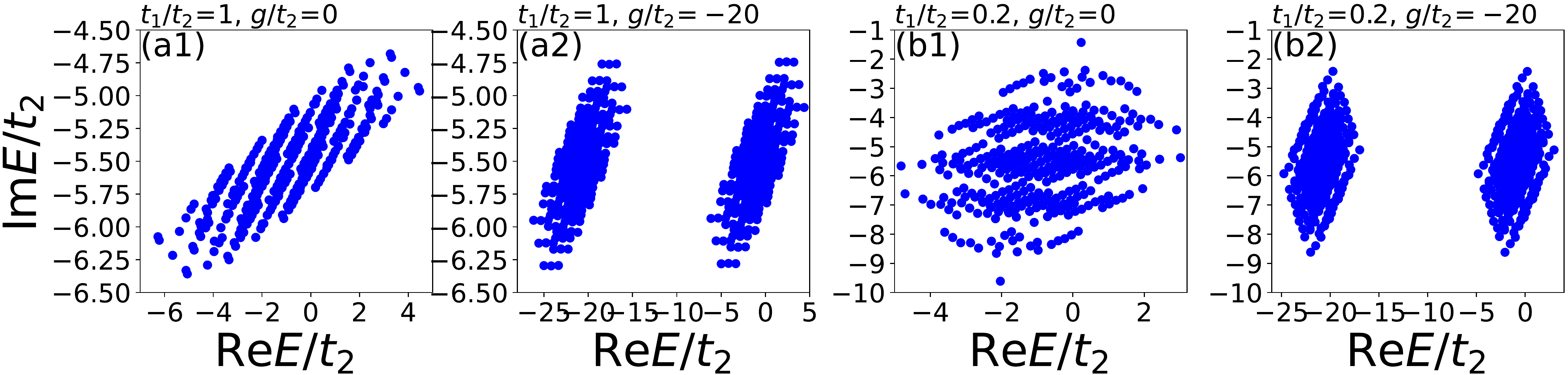}
\caption{Energy spectra for Hamiltonian (\ref{eq:H_SSH}) under OBCs with (a1): $t_{1}/t_{2}=1$ and $g=0$ (a2): $t_{1}/t_{2}=1$ and $g/t_2=-20$. (b1): $t_{1}/t_{2}=0.2$ and $g=0$. (b2) $t_{1}/t_{2}=0.2$ and $g/t_2=-20$. The other parameters are $t_2=(2\pi)\times1000$ Hz~\cite{Atala2013np}, the impurity position is $x_0=1$ ($A$ site), and $\tilde{\gamma}/t_{2}=0.92-0.15i$ at half filling $N=6$ and $2L=12$. Noticeably, the effect of the local interaction $g$ is non-local, drastically affecting the entire spectrum whether under OBCs (shown here) or PBCs (see main text).}
\label{fig:energy_OBC}
\end{figure}

Our interacting Hamiltonian is given by
\begin{align}\label{eq:H_SSH}
\hat{H}&=\sum_{x=1}^{L}\left[ - i\tilde{\gamma}(\hat{c}_{j,A}^{\dagger}\hat{c}_{j,A} + \hat{c}_{j,B}^{\dagger}\hat{c}_{j,B}) + \left(t_{1}+\tilde{\gamma}\right)\hat{c}_{x,A}^{\dagger}\hat{c}_{x,B} + \left(t_{1}-\tilde{\gamma}\right)\hat{c}_{x,B}^{\dagger}\hat{c}_{x,A}\right] \nonumber\\
&~~+ t_{2}\sum_{x=1}^{L-1}(\hat{c}_{x+1,A}^{\dagger}\hat{c}_{x,B} + {\rm H.c.}) + g\hat{n}_{x_0,s}^{(b)}\hat{n}_{x_0,s},
\end{align}
where $\hat{n}_{x_0,s}=\hat{c}_{x_0,s}^{\dagger}\hat{c}_{x_0,s}$ is the number operator at site $s$ of cell $x_0$ for environment atoms. $\hat{n}_{x_0,s}^{(b)}=\hat{b}_{x_0,s}^{\dagger}\hat{b}_{x_0,s}$ is the number operator at site $s$ of cell $x_0$ for impurity atom.
In our exact diagonalization calculations, we consider the system with half filling condition (see Fig.~\ref{fig:energy_OBC}).

To compare with the energy spectra under PBC in the main text, we plot the energy spectra under OBC as shown in Fig.~\ref{fig:energy_OBC}. It is found that the impurity interaction can also non-perturbatively split the entire OBC energy spectrum into two halves as those under PBC.

\newpage
\subsection{Spatial density}

\subsubsection{Spatial density of the long-time steady state}

In the fock space basis $\hat{F}=(\langle1|,\langle2|,...,\langle C_{2L}^{N}|)^{T}$, the number density operator at site $x$ is given by
\begin{align}
\hat{n}_{\text{fock}}(x)=\hat{F}\hat{n}_{x}\hat{F}^{\dagger}=\left(
  \begin{array}{cccccc}
    \langle1|\hat{n}_{x}|1\rangle & \langle1|\hat{n}_{x}|2\rangle & \cdot\cdot\cdot & \langle1|\hat{n}_{x}|C_{2L}^{N}\rangle \\
    \langle2|\hat{n}_{x}|1\rangle & \langle2|\hat{n}_{x}|2\rangle & \cdot\cdot\cdot & \langle2|\hat{n}_{x}|C_{2L}^{N}\rangle \\
    \cdot\cdot\cdot & \cdot\cdot\cdot & \cdot\cdot\cdot & \cdot\cdot\cdot \\
    \langle C_{2L}^{N}|\hat{n}_{x}|1\rangle & \langle C_{2L}^{N}|\hat{n}_{x}|2\rangle & \cdot\cdot\cdot & \langle C_{2L}^{N}|\hat{n}_{x}|C_{2L}^{N}\rangle
  \end{array}
\right),
\end{align} where $\hat{n}_{x}=\hat{c}_{x}^{\dagger}\hat{c}_{x}$.

\begin{figure}[h]
\includegraphics[width=.9\linewidth]{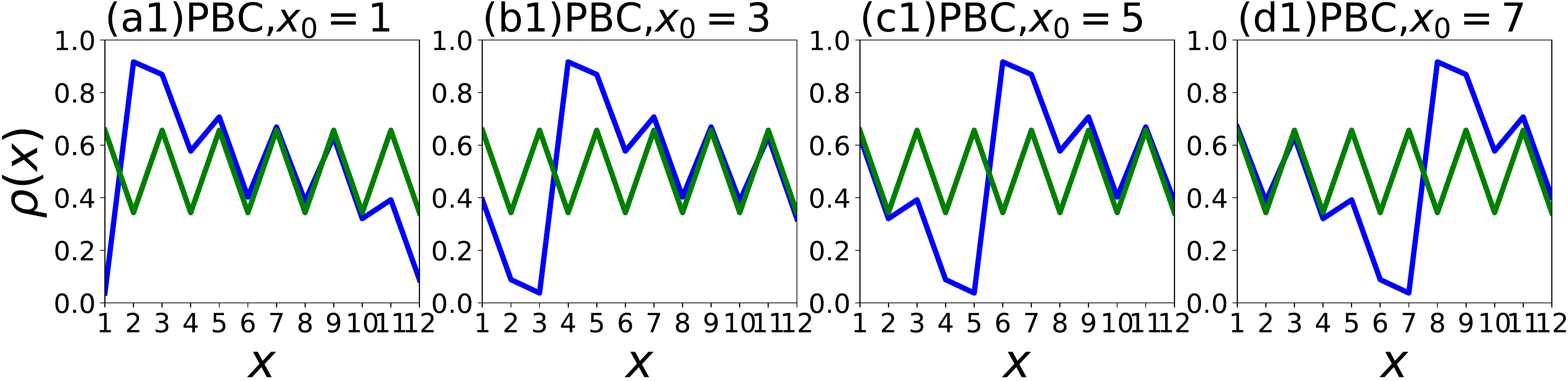}
\includegraphics[width=.9\linewidth]{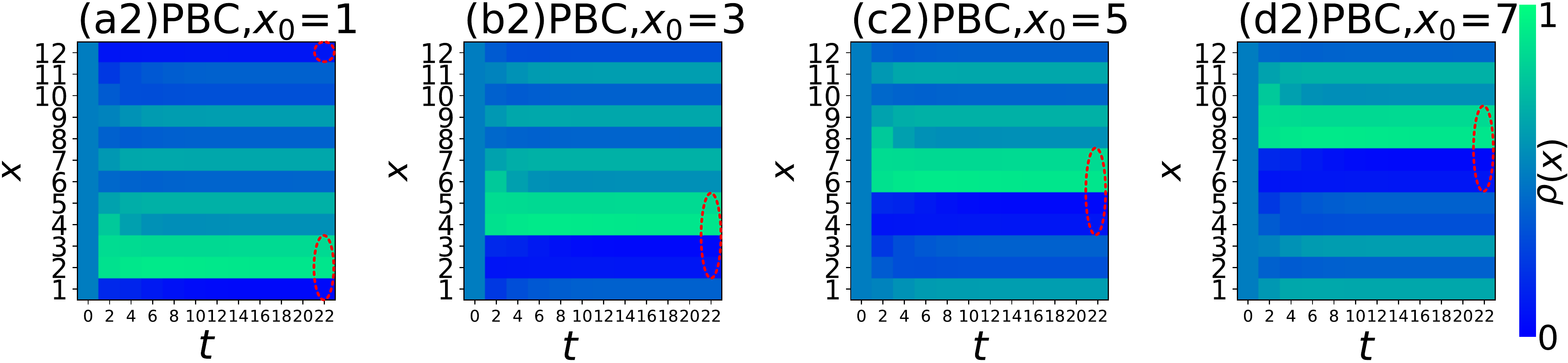}
\includegraphics[width=.9\linewidth]{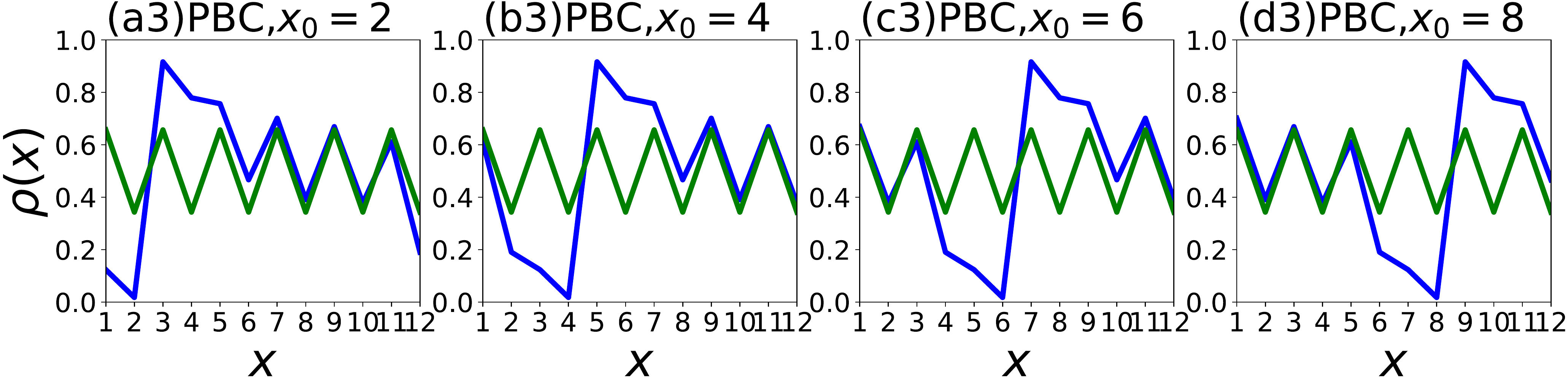}
\includegraphics[width=.9\linewidth]{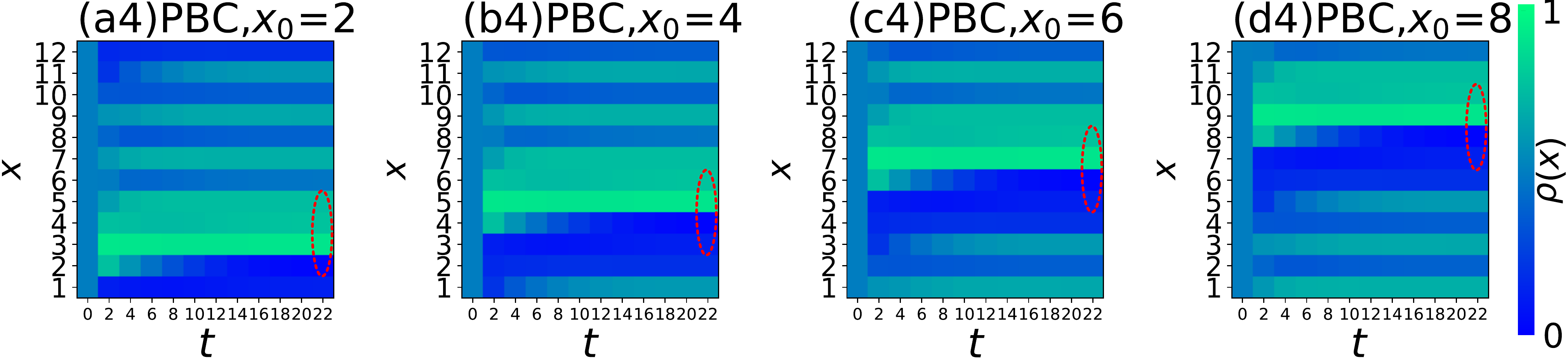}
\caption{(a1-d1) Long-time steady-state spatial density in Eq.~(\ref{eq:rho}) under PBCs for different odd impurity positions $x_0=1,3,5,7$ respectively. Note its translation invariant profile, which is exhibits the characteristic antisymmetric squeezed polaron profile (blue) for $g/t_{2}=-10$, or reference non-interacting case (green) with $g=0$. Parameters are $t_1/t_2=1$ at half filling $N=6$ and $2L=12$, $\tilde{\gamma}/t_{2}=0.92-0.15i$, $t_2=(2\pi)\times1000$ Hz~\cite{Atala2013np}, and the evolution time is in units of $t_2$.
(a2-d2) Dynamics of the spatial density along the lattice for different odd impurity positions $x_0=1,3,5,7$ respectively with $g/t_{2}=-10$.
(a3-d3) Long-time steady-state spatial density in Eq.~(\ref{eq:rho}) under PBCs for different even impurity positions $x_0=2,4,6,8$ respectively.
(a4-d4) Dynamics of the spatial density along the lattice for different even impurity positions $x_0=2,4,6,8$ respectively with $g/t_{2}=-10$.
Here, the initial state is $\left|\psi^{R}(0)\right\rangle=(\left|101010101010\right\rangle+\left|010101010101\right\rangle)/\sqrt{2}$. The antisymmetric dipole-like squeezed polaron (red circled) emerges after a short time, and persists in the steady state.}
\label{fig:density_PBC}
\end{figure}

\begin{figure}[h]
\includegraphics[width=.9\linewidth]{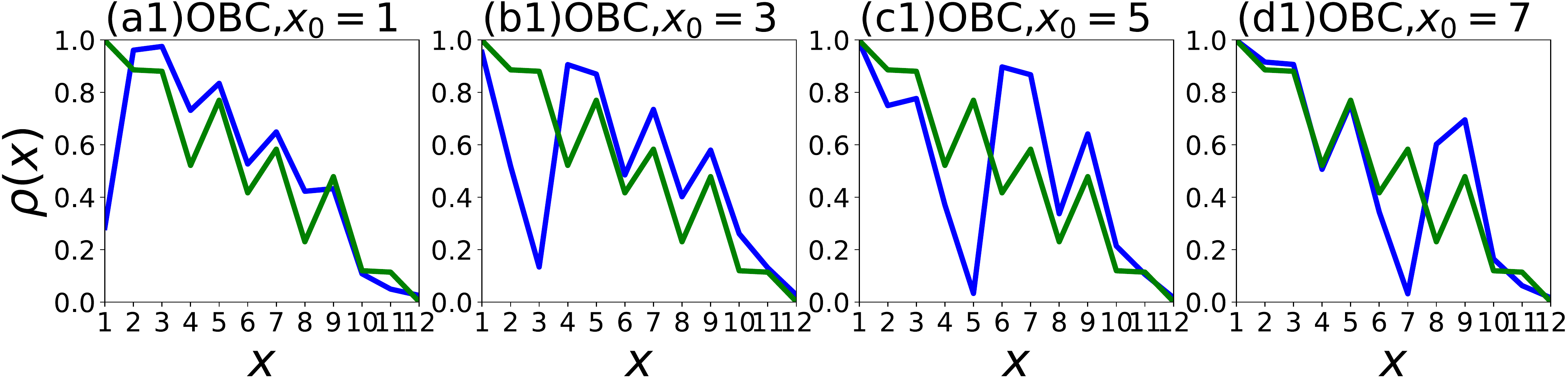}
\includegraphics[width=.9\linewidth]{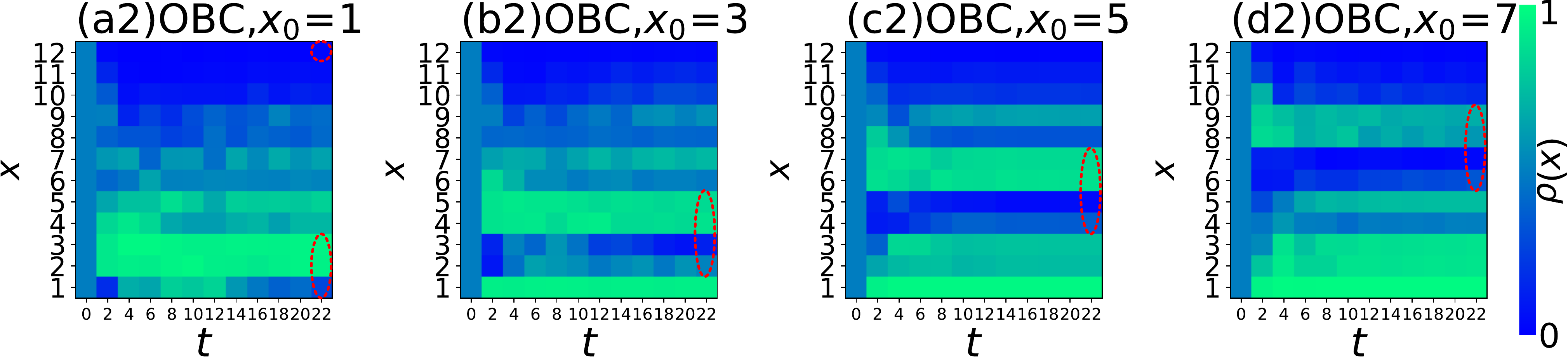}
\includegraphics[width=.9\linewidth]{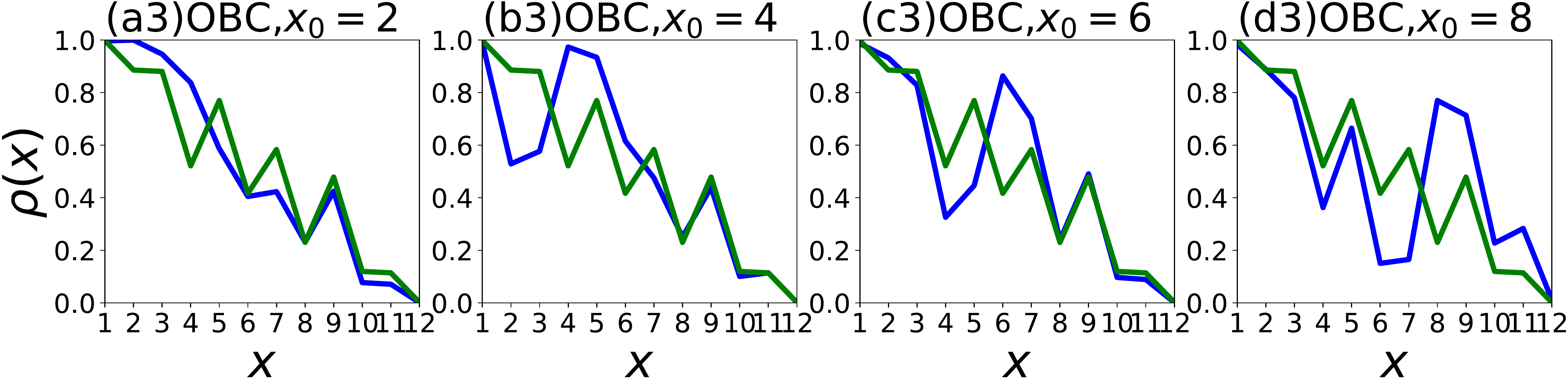}
\includegraphics[width=.9\linewidth]{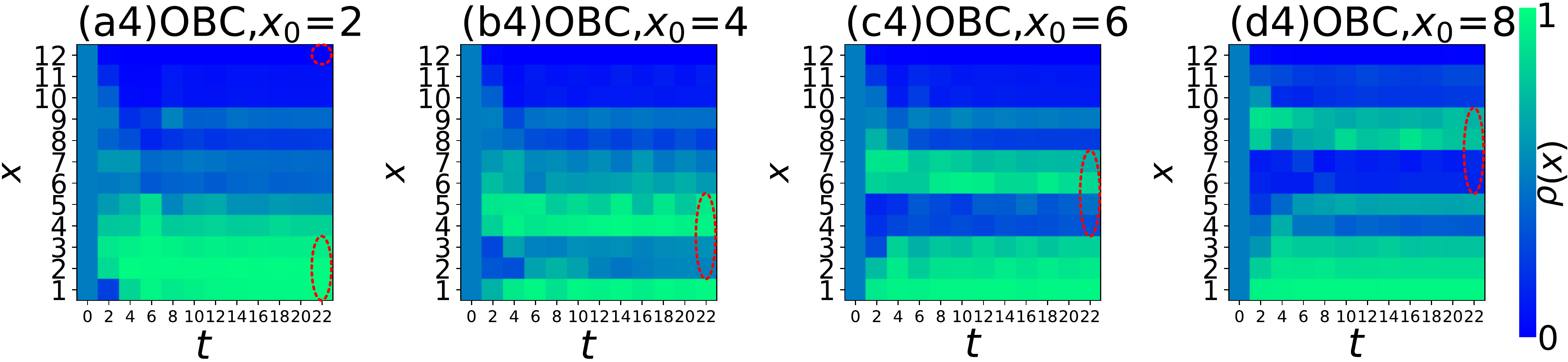}
\caption{(a1-d1) Long-time steady-state spatial density in Eq.~(\ref{eq:rho}) under OBCs for different odd impurity positions $x_0=1,3,5,7$ respectively.
(a2-d2) Dynamics of the spatial density along the lattice for different odd impurity positions $x_0=1,3,5,7$ respectively with $g/t_{2}=-10$.
(a3-d3) Long-time steady-state spatial density in Eq.~(\ref{eq:rho}) under OBCs for different even impurity positions $x_0=2,4,6,8$ respectively.
(a4-d4) Dynamics of the spatial density along the lattice for different even impurity positions $x_0=2,4,6,8$ respectively with $g/t_{2}=-10$.
The other parameters are the same with Fig.~\ref{fig:density_PBC}.
Note that even though finite-size effects become more prominent due to the boundaries, the characteristic antisymmetric dipole-like squeezed polaron profile (red circled) can still be seen, \emph{independent of the} inevitable non-Hermitian skin effect at the $x=1$ boundary. }
\label{fig:density_OBC}
\end{figure}

We define the spatial density of the long-time steady state (see the time evolutions in Figs.~\ref{fig:density_PBC} and \ref{fig:density_OBC})
\begin{align}\label{eq:rho}
	\rho(x)\equiv\lim_{t\to\infty}\langle\psi^{R}(t)|\hat{n}_{x}|\psi^{R}(t)\rangle,
\end{align}
where $|\psi^{R}(t)\rangle=e^{-i\hat{H}t}\left|\psi^{R}(0)\right\rangle/\left\|\left|e^{-i\hat{H}t}\psi^{R}(t)\right\rangle\right\|$ which is the state under quench dynamics of a prepared initial state $\left| \psi^{R}(0)\right\rangle=(\left|1010\cdots\right\rangle+\left|0101\cdots\right\rangle)/\sqrt{2}$ with the Hamiltonian $\hat{H}$. We choose the final state to be after a sufficiently long-time evolution such that the spatial density reaches a steady configuration.

To understand the steady state obtained under non-Hermitian time evolution, we consider an initial state expressed in eigenvalue basis $|\psi(0)\rangle=\sum_{n} c_{n}\left|E_{n}\right\rangle$, and express the time dynamics as
\begin{align}\label{dy1}
e^{-i\hat{H}t}|\psi(0)\rangle=\sum_{n} e^{-ic_{n}{\rm Re}(E_{n})t+c_{n}{\rm Im}(E_{n})t}\left|E_{n}\right\rangle.
\end{align}
Thus, long-time dynamics will converge to the eigenstate with the largest imaginary part of the eigenvalue:
\begin{align}\label{eq:dy}
	\lim_{t\to\infty}e^{-i\hat{H}t}|\psi(0)\rangle\rightarrow \left|{\rm Max }\,{\rm Im}(E_{n})\right\rangle.
\end{align}

Comparing Figs.~\ref{fig:density_PBC} and \ref{fig:density_OBC}, one can find that in the even-site impurity position cases, there is one-site shift for the squeezed polaron between PBC and OBC. But in the odd-site impurity position cases, there is no such shift for the squeezed polaron between PBC and OBC. This is a sublattice effect. Furthermore, Fig.~\ref{fig:density_OBC} shows that even though finite-size effects become more prominent due to the boundaries, the characteristic antisymmetric dipole-like squeezed polaron profile (red circled) can still be seen, \emph{independent of the} inevitable non-Hermitian skin effect at the $x=1$ boundary.

\clearpage
\subsubsection{Spatial density in a dynamical quench}

In order to investigate the necessity of the confluence of non-Hermiticity, non-reciprocity, and impurity interaction for observing squeezed polarons, we consider quench dynamics on spatial density with different impurity interaction strengths in our cold-atom experimental proposal model Hamiltonian (Eq. (5) in the main text). We activate the dissipation at a specific time:
\begin{equation}\label{eq:quenching}
\hat{H}(t)=
\begin{cases}
\hat{H}_{\tilde{\gamma}=0}, & (0\leqslant t < t_{c}); \\
\hat{H}_{\tilde{\gamma}\neq 0}, & (t\geqslant t_{c}),
\end{cases}
\end{equation} where $t_{c}$ is the time when the quench is added, and $\hat{H}$ is the Hamiltonian in Eq.~(\ref{eq:H_SSH}).
	
\begin{figure}[h]
\centering
\includegraphics[width=0.9\linewidth]{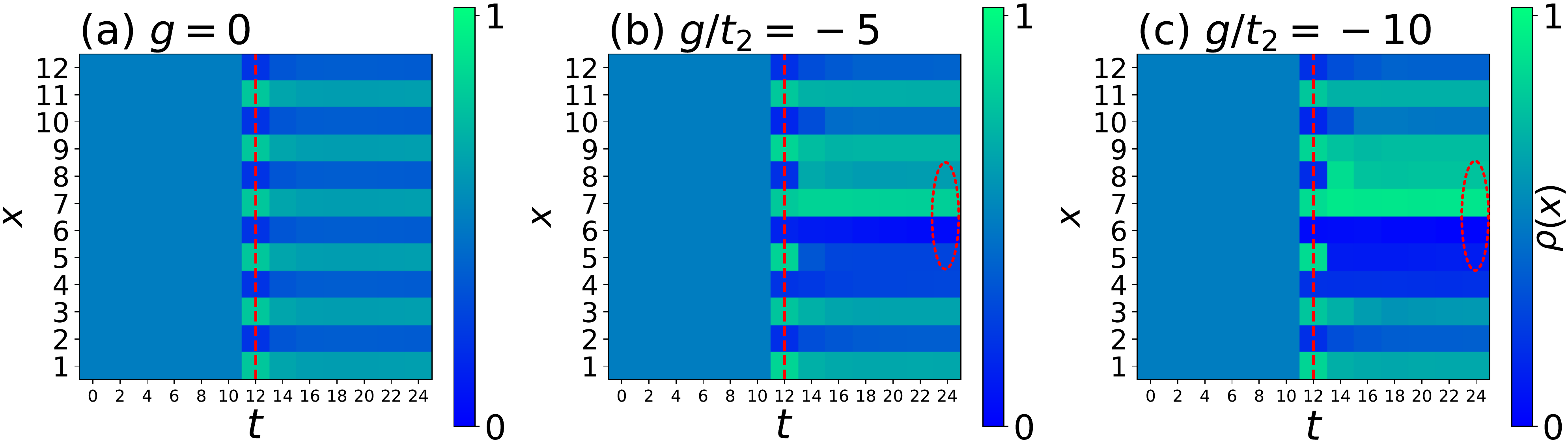}
\caption{Quench dynamics according to Eq.~(\ref{eq:quenching}) under PBCs such that we activate the dissipative loss at $t_{c}$ (red dashed line) for (a) $g=0$, (b) $g/t_2=-5$, and (c) $g/t_2=-10$. We set the impurity position $x_0=6$ ($B$ site), $\tilde{\gamma}/t_2=0.92-0.15i$, $t_1/t_2=1$ with $N=6$ fermions in $2L=12$ sites, and prepare the initial state as $\ket{\psi^{R}(0)}=(\ket{101010101010}+\ket{010101010101})/\sqrt{2}$. Evidently, the antisymmetric dipole-like squeezed polaron profiles (red circled) do not appear until after the non-Hermitian dissipation is switched on, even for the case with nonzero $g$. This shows that \underline{Hermitian polarons do not manifest in the time-evolution} of the given initial state (and in fact most other initial states that are not ground states), unlike non-Hermitian polarons which \underline{universally appear as a long-time steady-state property}. For $g=0$, there is spatial density only alternates between even and odd sites, without assuming any distinct antisymmetric profile.}\label{fig:quenching}
\end{figure}

As shown in the above three Figs.~\ref{fig:quenching}(a,b,c), for $0<t<t_{c}$ without the non-Hermiticity and non-reciprocity, the spatial density always equals to $1/2$ which means that there is no squeezed polarons. Besides, it indicates from Fig.~\ref{fig:quenching}(a) that the squeezed polarons also cannot be formed without impurity interaction. This demonstrates a very important point: while Hermitian polarons only manifest in the ground state, non-Hermitian polaron behavior, i.e., asymmetric dipole-like accumulation is much more universal, appearing \emph{almost universally} in the long-time steady-state behavior.

Indeed, our squeezed polarons are interesting dipole-like accumulations of fermionic density that only exist when all the three ingredients: non-Hermiticity, non-reciprocity, and impurity interaction, are simultaneously present.

Besides, we can also consider the quench dynamics with impurity interaction suddenly being turned on at a specific time $t_c$:
\begin{equation}\label{eq:quenching_g}
\hat{H}(t)=
\begin{cases}
\hat{H}_{g=0}, & (0\leqslant t< t_{c}); \\
\hat{H}_{g\neq 0}, & (t\geqslant t_{c}),
\end{cases}
\end{equation} where $\hat{H}$ is the Hamiltonian in Eq.~(\ref{eq:H_SSH}).

\begin{figure}[h]
\centering
\includegraphics[width=0.6\linewidth]{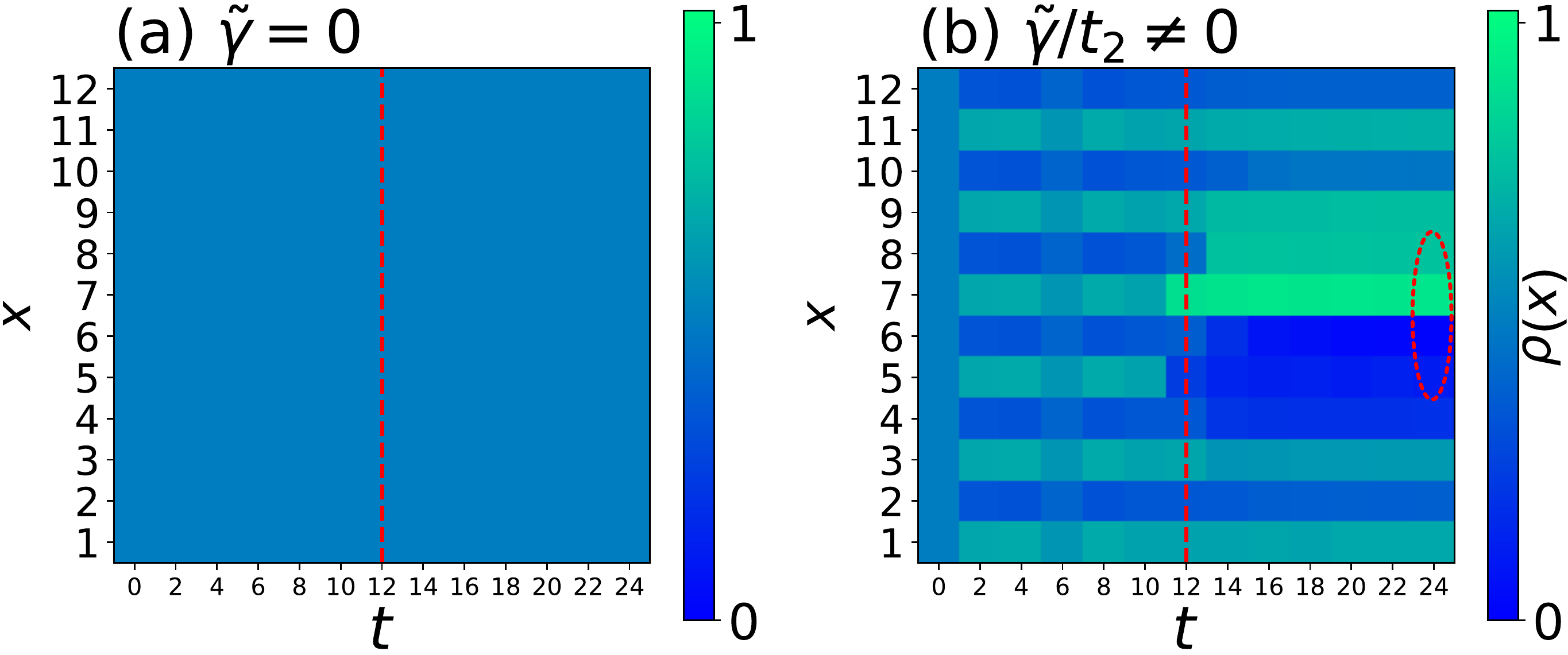}
\caption{Quench dynamics in Eq.~(\ref{eq:quenching_g}) under PBCs such that we activate the impurity interaction at $t_{c}$ (red dashed line) for (a) $\tilde{\gamma}=0$ and (b) $\tilde{\gamma}/t_2=0.92-0.15i$. We set the impurity position $x_0=6$ ($B$ site), $g/t_2=-10$, $t_1/t_2=1$ with $N=6$ fermions in $2L=12$ sites, and prepare the initial state as $\ket{\psi^{R}(0)}=(\ket{101010101010}+\ket{010101010101})/\sqrt{2}$. Evidently, the squeezed polaron (red circled) only exists for $\tilde\gamma\neq 0$, and also after the impurity interaction $g$ is turned on.}\label{fig:quenching_g}
\end{figure}

As shown in Fig.~\ref{fig:quenching_g}(a), for the Hermitian case, no matter adding impurity interaction or not, the spatial density always equals to $1/2$ which means that there is no squeezed polarons. Besides, it is indicated from Fig.~\ref{fig:quenching_g}(b) that, for the non-Hermitian case, the squeezed polarons also cannot be formed without impurity interaction. Therefore, our squeezed polarons only exist when all the three ingredients: non-Hermiticity, non-reciprocity, and impurity interaction, are simultaneously present.

\end{document}